\documentclass{aa}

\usepackage{amssymb}
\usepackage{amsmath}
\usepackage{graphicx}
\usepackage{txfonts}
\usepackage{booktabs}

\usepackage{textcomp}
\bibpunct{(}{)}{;}{a}{}{,}

\begin{document}

\title{False outliers of the $E_{\mathrm{p,i}}$ -- $E_{\mathrm{iso}}$ correlation?}
\titlerunning{Amati out}
\authorrunning{R. Martone et al.}
\author{R. Martone\textsuperscript{1,2}, L. Izzo\textsuperscript{3}, M. Della Valle\textsuperscript{4,5}, L. Amati\textsuperscript{6}, G. Longo\textsuperscript{2}, D. G\"otz\textsuperscript{7}.}
\institute{\textsuperscript{1}Dipartimento di Fisica e Scienze della Terra, Universit\'a di Ferrara, via Saragat 1, I-44122, Ferrara, Italy\\
\textsuperscript{2}Universit\'a di Napoli "Federico II", Dipartimento di Fisica "Ettore Pancini", C. U. Monte S. Angelo, 80126 Napoli, Italy\\
\textsuperscript{3}Instituto de Astrofisica de Andalucia (IAA-CSIC), Glorieta de la Astronomia s/n, 18008 Granada, Spain \\
 \textsuperscript{4}INAF-Osservatorio Astronomico di Napoli, Salita Moiariello, 16, I-80131 Napoli, Italy \\ 
 \textsuperscript{5}International Center for Relativistic Astrophysics, Piazza della Repubblica 10, I-65122 Pescara, Italy \\  
 \textsuperscript{6}INAF-IASF Bologna, via P. Gobetti, 101. I-40129 Bologna, Italy \\
 \textsuperscript{7}AIM--CEA/DRF/Irfu/Service d'Astrophysique, Orme des Merisiers, F-91191 Gif-sur-Yvette, France}

\abstract{In the context of an in-depth understanding of GRBs and their possible use in cosmology, some important correlations between the parameters that describe their emission have been discovered, among which the "$E_{\mathrm{p,i}}$ -- $E_{\mathrm{iso}}$" correlation is the most studied. Because of this, it is fundamental to shed light on the peculiar behaviour of a few events, namely GRB 980425 and GRB 031203, that appear to be important outliers of the $E_{\mathrm{p,i}}$ -- $E_{\mathrm{iso}}$ correlation.}{In this paper we investigate if the locations of GRB 980425 and GRB 031203, the two (apparent) outliers of the correlation, may be due to an observational bias caused by the lacking detection of the soft X-ray emissions associated with these GRBs, from respectively the Burst And Transient Source Experiment (BATSE) detector on-board the Compton Gamma-Ray Observer and INTEGRAL, that were operating at the epoch at which the observations were carried out. We analyse the observed emission of other similar sub-energetic bursts (GRBs 060218, 100316D and 161219B) observed by Swift and whose integrated emissions match the $E_{\mathrm{p,i}}$ -- $E_{\mathrm{iso}}$ relation. We simulate their integrated and time-resolved emissions as would have been observed by the same detectors that observed GRB 980425 and GRB 031203, aimed at reconstructing the light curve and spectra of these bursts.} {We estimate the $E_{\mathrm{p,i}}$ and the $E_{\mathrm{iso}}$ parameters from the time-resolved and total integrated simulated spectra of GRBs 060218, 100316D and 161219B as observed by BeppoSAX, BATSE, INTEGRAL and the WFM proposed for the LOFT \citep{LOFT2012} and eXTP missions \citep{Zhang2016}.}{If observed by old generation instruments, GRB 060218, 100316D and 161219B would appear as outliers of the $E_{\mathrm{p,i}}$--$E_{\mathrm{iso}}$ relation, while if observed with Swift or WFM GRB 060218 would perfectly match the correlation. We also note that the instrument BAT alone (15-150 keV) actually measured 060218 as an outlier.}
{We suggest that if GRB 980425 and GRB 031203 would have been observed by Swift and by eXTP they may have matched the $E_{\mathrm{p,i}}$--$E_{\mathrm{iso}}$ relation. This provides strong support to the idea that instrumental biases can make some events in the lower-left corner of the $E_{\mathrm{p,i}}$ -- $E_{\mathrm{iso}}$ plane appearing as outliers of the "Amati relation".}


\date{}

\maketitle

\section{Introduction}\label{sec:1}

Observations of long duration gamma-ray bursts (GRBs) in the last decades have pointed out the existence of a large 
number of empirical relations which link some of the fundamental parameters of GRBs such as, for example, the isotropic 
energy $E_{\mathrm{iso}}$ emitted in gamma rays, the peak energy of the prompt emission spectrum $E_{\mathrm{p,i}}$, the peak luminosity 
$L_{\mathrm{p}}$ of the prompt emission 
\citep{Amati2002,Ghirlanda2004,Yonetoku2004,LiangZhang,Dainotti2008,Bernardini2012,Margutti2013,Izzo2015}.

In this work we focus on the most popular of them, the $E_{\mathrm{p,i}}$ -- $E_{\mathrm{iso}}$ correlation, aka the 'Amati 
relation' \citep{Amati2002, Amati2006}: the total gamma-ray isotropic energy ($E_{\mathrm{iso}}$) emitted in long GRBs correlates 
with the rest-frame value of the energy spectrum at which their gamma-ray emission peak ($E_{\mathrm{p,i}}$). In this paper the 
isotropic output is extimated using the quantity $E_{\mathrm{\gamma, iso}}$ that represents the total energetic output in the 
rest-frame range 1-10000 keV.

To date more than 200 GRBs match the $E_{\mathrm{p,i}}$--$E_{\mathrm{iso}}$ relation; however, after 20 years, it is still a matter of 
debate the fact that the closest GRBs ever discovered, GRB 980425 at z=0.0085 (d = 40 Mpc), appears to be a remarkable 
outlier of the 'Amati relation' \citep{Ghisellini2006, Amati2006}. This situation is still more disturbing after noting 
that GRB 980425 was found to be the first GRB associated with a Supernova (SN), the SN 1998bw \citep{Galama1998}, and 
therefore it is recognized as the prototype of GRB-SN connection \citep{Woosley2006,DellaValle2011}. The existence of 
outliers of the 'Amati relation' should be also clarified in view of both understanding the emission processes at play 
in the GRB phenomenon and the frequent use of GRBs in cosmological studies \citep{Amati2008, cosmo2013, Izzo2015}. In 
this paper we suggest that the location of GRB 980425 in the $E_{\mathrm{p,i}}$ -- $E_{\mathrm{iso}}$ plane is very likely due to an 
observational bias caused by the sensitivity range (25-2000) keV of the Burst And Transient Source Experiment (BATSE) 
detector on-board the Compton Gamma-Ray Observer (CGRO, \citealp{Meegan1992}). Similar arguments apply to the case of an 
other sub-energetic and nearby (z = 0.105) event: GRB 031203 \citep{Mazzali2006, Watson20062}.

\begin{figure}
\centering
\includegraphics[scale=0.50]{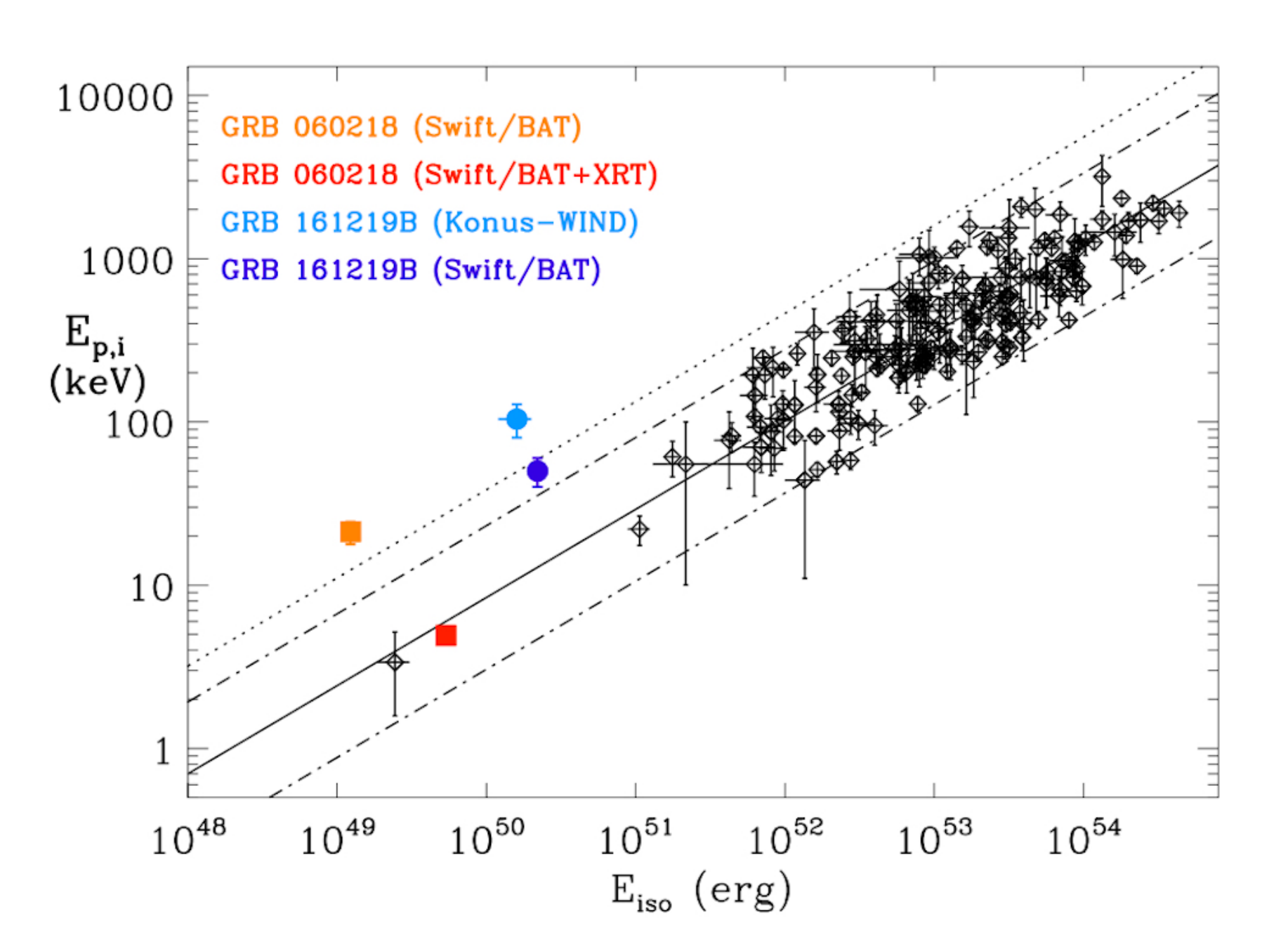}\\
\caption{Location in the $E_{\mathrm{p,i}}$ -- $E_{\mathrm{iso}}$ plane of GRB\,161219B as observed by Swift/BAT and Konus-WIND 
and of GRB\,060218 as observed by Swift/BAT and by Swift/BAT+XRT. Swift/BAT 
is more sensitive than Konus-WIND, thus allowing a more precise estimate of the $E_{\mathrm{p,i}}$ and 
$E_{\mathrm{iso}}$ parameters for GRB 161219B and findig it more consistent with the Amati relation. In the outstanding case of 
GRB\,060218, the emission in the soft X-ray band, that can be detected only by using Swift/XRT, makes this event, which 
otherwise would have been classified as an outlier, fully consistent with the $E_{\mathrm{p,i}}$ -- $E_{\mathrm{iso}}$ correlation. In 
the plot, 
the dotted-dashed (dotted) lines refer to the 2 (3) sigma error around the best-fit 
line.}
\label{fig:Lorenz}
\end{figure}

\begin{figure}
\centering
\includegraphics[scale=0.30]{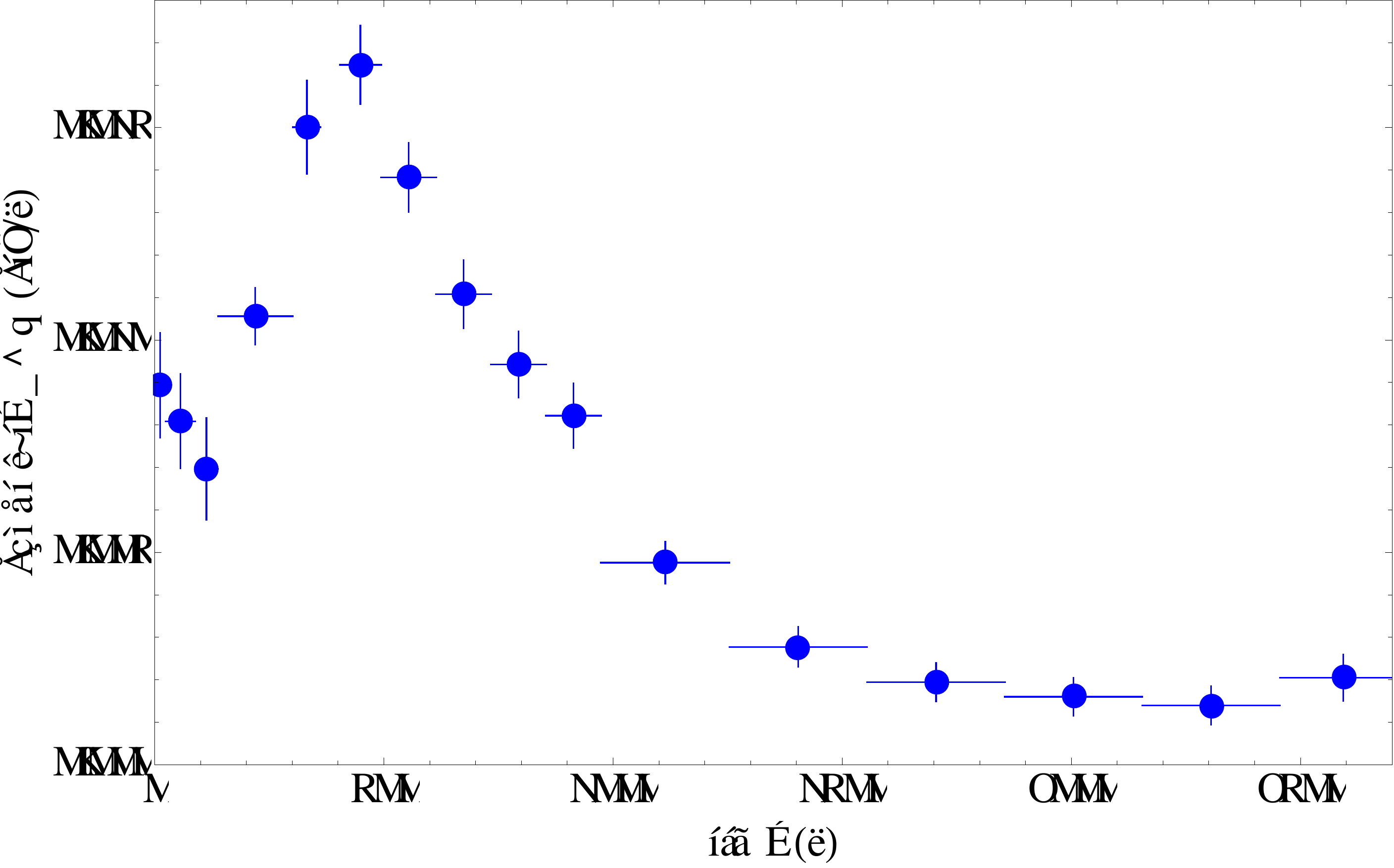}
\includegraphics[scale=0.30]{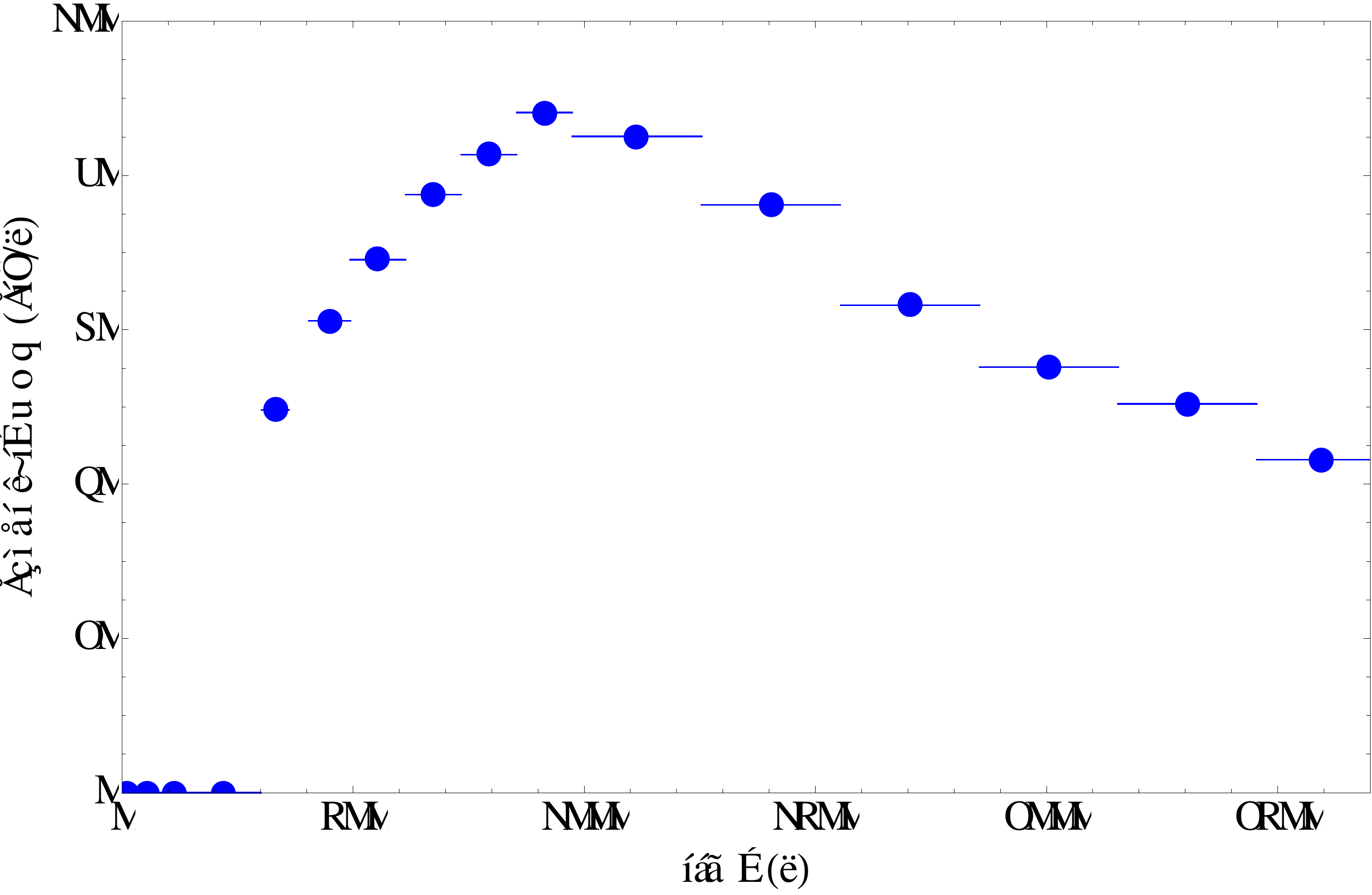}
\includegraphics[scale=0.30]{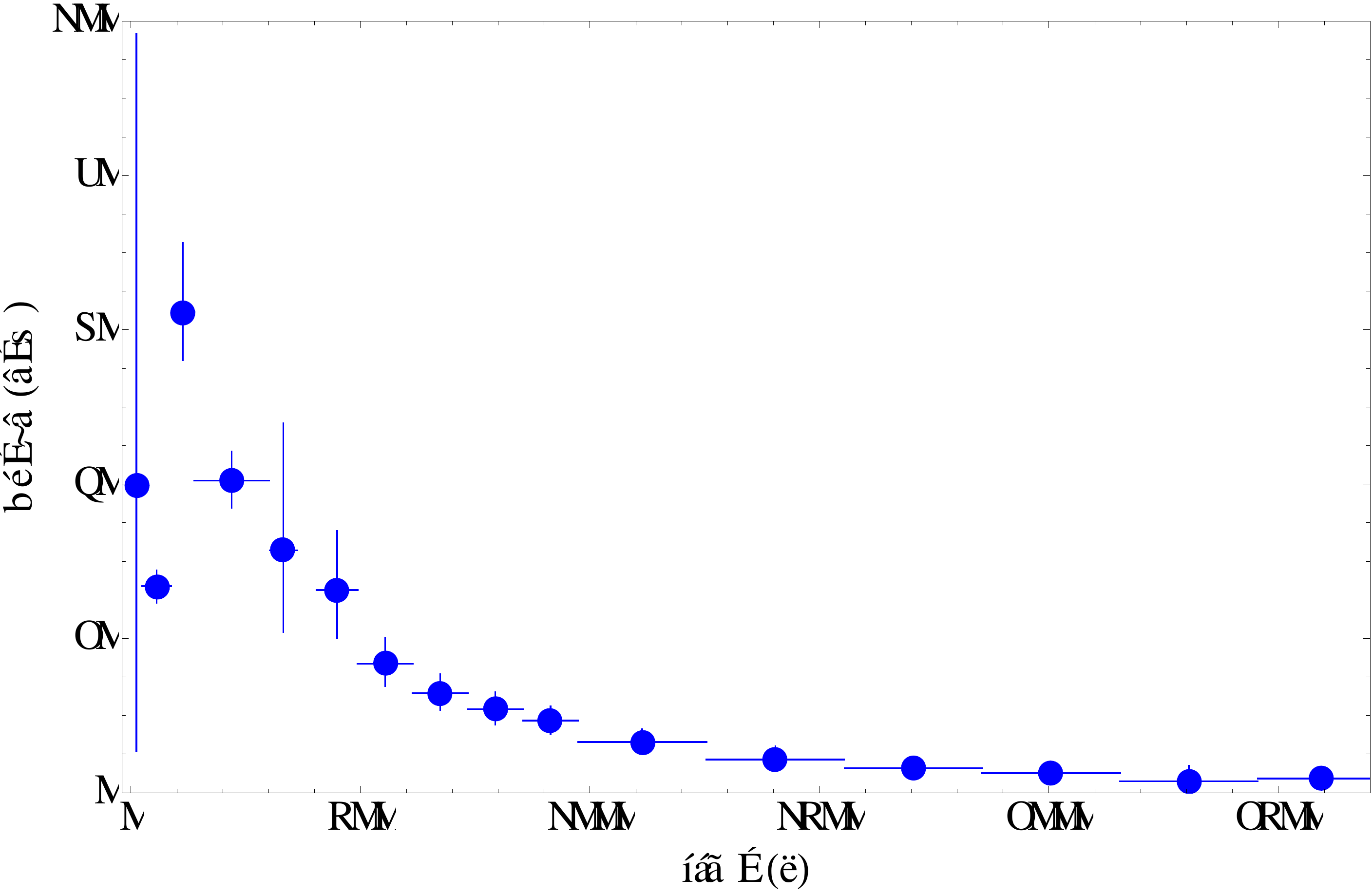}\\
\caption{\textbf{(upper panel)} The net count rate as detected by Swift BAT (15-150 keV) and \textbf{(middle panel)} by Swift XRT (0.3-10 keV) after pile-up correction. \textbf{(lower panel)} The variation of the intrinsic peak energy of GRB 060218 as detected by Swift.}
\label{fig:no1}
\end{figure}

To reach our goal, we show that nearby and sub-energetic bursts with a SN associated, GRB 060218 \citep{Campana2006}, 
GRB 100316D \citep{Starling2011} and GRB 161219B \citep{Cano2017} observed by the Swift Burst Alert Telescope (BAT, 
\citealp{Barthelmy2005}) in the energy range (15-150) keV and the X-Ray Telescope (XRT, \citealp{Burrows2005}) 
in the energy range (0.3-10) keV, consistent with the $E_{\mathrm{p,i}}$ -- $E_{\mathrm{iso}}$ relation,
would appear as outliers of the Amati relation if observed with BATSE. 

These GRBs are perfect for our purposes because,
unlike other similar low-energetic events, they have
a continuous coverage in time of their prompt emission by
Swift BAT and, in the case of GRB\,060218 and GRB\,100316D, also by XRT.
The importance of the different instruments
characteristics in determining the position of an event
in the $E_{\mathrm{p,i}}$ -- $E_{\mathrm{iso}}$ plane can be appreciated considering figure \ref{fig:Lorenz},
where we highlight the position of GRB\,060218 and GRB\,161219B according to
different detectors: it is clearly
visible that when using measurements by instruments with better sensitivity and lower energy threshold these events 
become more consistent with the correlation.
GRB\,060218 is the emblem of this kind of behaviour, perfectly matching the best--fit of the $E_{\mathrm{p,i}}$ -- $E_{\mathrm{iso}}$ 
correlation when seen by 
Swift/XRT+BAT, and appearing as an outlier when observed with wift/BAT only, as will be shown in this work. 

This work is organized as follows: in section 2 we present the spectral properties of GRBs 060218, 100316D and 
161219B and we introduce the methodology at the base of this paper. In Sec. 3 we describe their spectral analysis and in 
Sec. 4 we present the simulations of these GRBs as if they were observed by BATSE and other detectors. In the last 
section we report our conclusions.

\section{Swift data analysis}

In the following part of this article, we will mainly focus on the case of GRB 060218, which presents one of the best dataset among the observed GRBs. Additional material regarding GRB 100316D and GRB 161219B, as figures and tables, can be found in the Appendix.

\subsection{GRB 060218}

GRB 060218 was discovered by Swift \citep{Campana2006} and it was found to be associated with SN 2006aj \citep{Pian2006} at the redshift of $z = 0.0331$. Soft X-ray observations pointed out the presence of a thermal component, which originated in the breakout of a shock propagating into the wind surrounding the progenitor star \citep{Campana2006, Waxman2007}. The main feature which makes this GRB different from more energetic GRBs is the long duration ($\sim$ 3000 s) of the prompt emission observed down to X-rays, which is clearly different from the canonical one observed in almost all GRBs \citep{Nousek2006}. Thank to this very long duration (and its proximity)  it was possible to detect most of the prompt emission with both BAT (15-150 keV) and XRT (0.2-10 keV). The integrated BAT+XRT spectrum is characterized by an intrinsic peak energy of $E_{\mathrm{p,i}} = 4.9$ keV and a total integrated isotropic energy of $E_{\mathrm{iso}} = 6.2 \times 10^{49}$ erg. With these values GRB 060218 matches the Amati relation (see Fig. 6).

Due to its low luminosity, low redshift and associated Supernova, GRB 060218 has been considered a ''twin'' of GRB 980425 and GRB 031203 \citep{Ghisellini2006}, but it shows a different time duration and high energy emission. It is consequently very interesting to derive the spectrum of GRB 060218 and its location in the $E_{\mathrm{p,i}}$ -- $E_{\mathrm{iso}}$ plane as it would have been observed by the same instruments which have observed GRB 980425, (BeppoSAX, \citep{Frontera2000}), BATSE, \citep{Meegan1992}) and GRB 031203 (INTEGRAL \citep{Mereghetti2003}). We also consider the case for eXTP \citep{Amati2013, Zhang2016}, a future-planned mission dedicated to observe the X-ray transient sky in the soft X-ray energies. 

We have reproduced the Swift data analysis as reported in \citet{Campana2006} using the same time intervals, and the results are reported in Fig. \ref{fig:no1}. The XRT spectral data were obtained for the corresponding BAT time intervals following the canonical procedure for GRB data reduction, starting with the \texttt{xrtpipeline} package, which runs in sequence all the tasks for XRT data processing. Since the X-ray emission from GRB 060218 was very bright, we have applied the pile-up correction for the Window Timing mode, as the source presented count rates larger than 100 counts s$^{-1}$ for a large part of its emission. In this light, we have selected a box with an annulus centered on the brightest pixel, as well described in  \citet{Romano2006}. After the pile-up correction, we have obtained background files with XSELECT and generated the corresponding ancillary response function file with the \texttt{xrtmkarf} package. Finally, we have grouped the data in order to have a minimum of ten counts in each spectral bin, using the \texttt{grppha} package. 

Since the complete dataset is composed of four spectra for which there are no XRT data, we have divided the sample in two sub-dataset: 1) the first four BAT spectra  lasting totally $t_{\mathrm{D1}} = 340$ s, and 2) the following twelve BAT and XRT spectra, lasting totally $t_{\mathrm{D2}} = 2387$ s and which cover the range (0.3 - 150) keV, with a data gap between (10 - 15) keV. The spectral data analysis has been performed using the XSPEC fitting package \citep{XSPEC2}, assuming Solar abundances as given in \citet{Wilms2000} and a cosmological model with $\Omega_{\mathrm{\Lambda}}$ = 0.73, $H_0$ = 70 km/s/Mpc and $q_0$ = -0.5. For the BAT + XRT dataset, we found that the best-fit in all single spectra is given by an absorbed black body plus a power-law with an exponential cut-off, in agreement with the results of \citet{Campana2006}. The results of the time-resolved spectral analysis of the two datasets are shown in Tables \ref{tab:no1}, \ref{tab:no2}, while in fig.2 we report as an example the best-fit for the Swift BAT+XRT spectrum number 5 using a function composed by an absorbed blackbody plus a power-law with an exponential energy cutoff.
\begin{figure}
\centering
\includegraphics[scale=0.32]{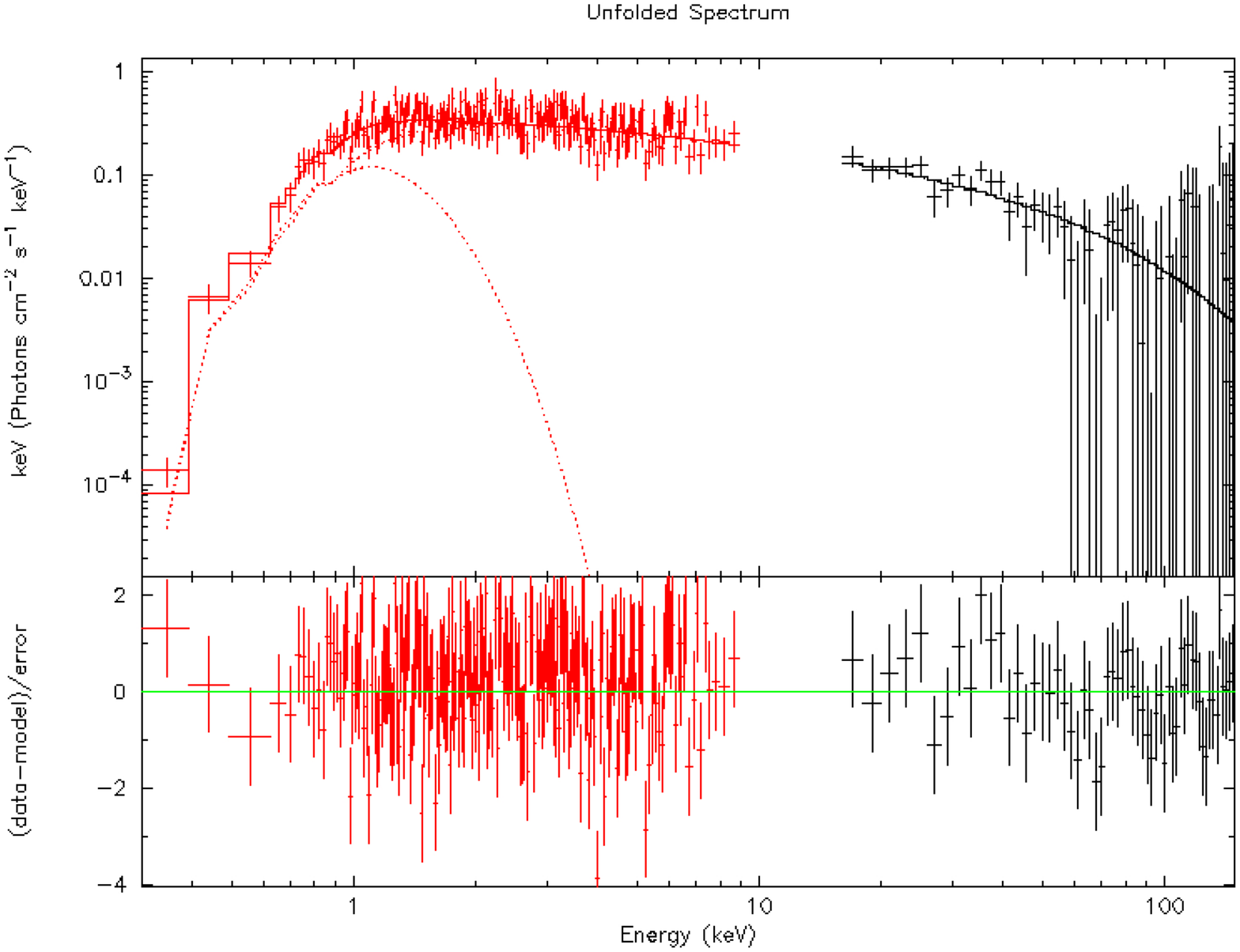}
\caption{Best-fit of the Swift BAT+XRT spectrum number 5 (see Table \ref{tab:no2}) obtained with an absorbed blackbody and a power-law with an exponential energy cutoff function.}
\label{Spectrum}
\end{figure} 

The last step consists in computing the integrated spectrum of GRB 060218. We have obtained integrated spectra for both datasets using the \texttt{mathpha} task, which is provided within the \texttt{heasoft} package for data analysis\footnote{http://heasarc.gsfc.nasa.gov/lheasoft}. We have then fitted the integrated spectra considering a cutoff powerlaw for the first dataset and an absorbed cutoff powerlaw \citep{Band2003} plus a blackbody for the second one, obtaining very similar results of the one presented in \citet{Campana2006}. We have fixed the galactic column density to the value $N(H_{\mathrm{gal}}) = 1.42 \times 10^{21}$ cm$^{-2}$, (see \citet{Dickey1990}), while for the extragalactic column density, we have chosen the median among those obtained in the different spectra in which our dataset is divided: $N(H_{\mathrm{intr}}) = 3.58 \times 10^{21}$ $cm^{-2}$. The results of the fits are shown in Table \ref{tab:no2}. From these results, we have computed the values for the intrinsic peak energy $E_{\mathrm{p,i}}$ and isotropic energy emitted in the time intervals corresponding to the two datasets, and the evolution of $E_{\mathrm{p}}$ is shown in Fig. \ref{fig:no1}. Finally, we compute the total integrated $E_{\mathrm{p,i}}$ and $E_{\mathrm{iso}}$ values for Swift, by considering a Band function alone and we obtain $E_{\mathrm{p,i,D2}}$ = 4.92 $\pm$ 0.57 keV and $E_{\mathrm{iso,D2}}$ = (3.10 $\pm$ 0.10) $\times$ 10$^{49}$ erg. 


\begin{table*}
\centering
\tiny{
\caption{Swift BAT (15-150 keV) spectral fits data results of the first dataset of GRB 060218, that includes the first four BAT spectra ($\Delta t_{\mathrm{D1}} = 340$ s).}
\label{tab:no1}
\begin{tabular}{l c c c c c c}     
\toprule
$\#$ & $\Delta$t  & $\gamma$ & $E_{\mathrm{cutoff}}$ & norm & flux & $\chi^2$/DOF\\
 & ( s ) & & ( keV ) &  Photons $keV^{-1}$ $cm^{-2}$ $s^{-1}$ & (10$^{-9}$ erg/cm$^2$/s) & \\
\midrule
1  &  64    &  $2.07_{\mathrm{-0.38}}^{+0.44}$ & - & $6.6_{\mathrm{-2.2}}^{+2.6} \times 10^{-3}$ & 5.0 & 63.2/56 \\
[0.15cm]
2  &  64  &  $2.61_{\mathrm{-0.37}}^{+0.43}$ & - & $9.6_{\mathrm{-2.5}}^{+2.9} \times 10^{-3}$ & 4.3 & 59.6/56 \\
[0.15cm]
3  &  49    &  $2.55_{\mathrm{-0.51}}^{+0.66}$ & - & $7.2_{\mathrm{-2.5}}^{+3.1} \times 10^{-3}$ & 3.4 & 43.0/56 \\
[0.15cm]
4  &  163   &  $0.91_{\mathrm{-1.07}}^{+0.85}$ & $35.9_{\mathrm{-3.0}}^{+3.2}$ & $9.9_{\mathrm{-0.95}}^{+0.95} \times 10^{-2}$ & 5.1 & 26.2/55 \\
\bottomrule
\end{tabular}}
\end{table*}

\begin{table*}
\centering
\tiny
\caption{Swift BAT+XRT (0.3-150 keV) spectral fits data results of the second dataset of GRB 060218, that includes the last 12 spectra ($\Delta t_{\mathrm{D2}} = 2387$ s).}
\label{tab:no2}
\begin{tabular}{l c c c c c c c c c}     
\toprule
$\#$ & $\Delta$t   & $\gamma$ & $E_{\mathrm{cutoff}}$  & norm CPO & kT & norm BB & flux BAT & flux XRT & $\chi^2$/DOF\\
& & & & & & & (10$^{-9}$) & (10$^{-9}$) &\\
& (s)	 & & (keV)  & Photons $keV^{-1}$ $cm^{-2}$ $s^{-1}$ & (keV)  & $10^{37}$ erg $s^{-1}$  & (erg/cm$^2$/s) & (erg/cm$^2$/s) &   \\
\midrule
5 & 60 & $1.36^{+0.11}_{\mathrm{-0.12}}$ & $47^{+23}_{\mathrm{-13}}$ & $0.51^{+0.07}_{\mathrm{-0.07}}$ & $0.197^{+0.033}_{\mathrm{-0.030}}$ & $1.23^{+0.30}_{\mathrm{-0.33}}$ & 7.0 & 3.7 & 277.8/281\\
[0.15cm]
6 & 90 & $1.361^{+0.067}_{\mathrm{-0.078}}$ & $39.7^{+10.8}_{\mathrm{-8.2}}$ & $0.68^{+0.05}_{\mathrm{-0.06}}$ & $0.171^{+0.023}_{\mathrm{-0.017}}$ & $1.19^{+0.34}_{\mathrm{-0.33}}$ & 7.7 & 4.7 & 391.9/412\\
[0.15cm]
7 & 120 & $1.290^{+0.065}_{\mathrm{-0.070}}$ & $22.8^{+4.1}_{\mathrm{-3.3}}$ & $0.80^{+0.05}_{\mathrm{-0.05}}$ & $0.151^{+0.018}_{\mathrm{-0.016}}$ & $1.67^{+0.47}_{\mathrm{-0.38}}$ & 5.5 & 5.5 & 398.3/505\\
[0.15cm]
8 & 120 & $1.159^{+0.084}_{\mathrm{-0.090}}$ & $14.9^{+2.4}_{\mathrm{-2.0}}$ & $0.80^{+0.06}_{\mathrm{-0.07}}$ & $0.177^{+0.017}_{\mathrm{-0.016}}$ & $2.05^{+0.35}_{\mathrm{-0.32}}$ & 4.1 & 6.1 & 491.6/527\\
[0.15cm]
9 & 120 & $1.244^{+0.082}_{\mathrm{-0.088}}$ & $13.9^{+2.4}_{\mathrm{-2.0}}$ & $0.96^{+0.07}_{\mathrm{-0.07}}$ & $0.169^{+0.017}_{\mathrm{-0.016}}$ & $2.12^{+0.40}_{\mathrm{-0.36}}$ & 3.3 & 6.2 & 553.9/539\\
[0.15cm]
10 & 120 & $1.225^{+0.082}_{\mathrm{-0.089}}$ & $11.7^{+2.0}_{\mathrm{-1.7}}$ & $1.06^{+0.07}_{\mathrm{-0.07}}$ & $0.162^{+0.015}_{\mathrm{-0.014}}$ & $2.47^{+0.41}_{\mathrm{-0.41}}$ & 2.7 & 6.6 & 461.5/542\\
[0.15cm]
11 & 280 & $1.296^{+0.065}_{\mathrm{-0.072}}$ & $9.0^{+1.3}_{\mathrm{-1.2}}$ & $1.19^{+0.05}_{\mathrm{-0.05}}$ & $0.150^{+0.008}_{\mathrm{-0.008}}$ & $2.98^{+0.37}_{\mathrm{-0.34}}$ & 1.3 & 6.0 & 717.9/670\\
[0.15cm]
12 & 300 & $1.15^{+0.18}_{\mathrm{-0.23}}$ & $4.9^{+1.7}_{\mathrm{-1.2}}$ & $1.15^{+0.05}_{\mathrm{-0.06}}$ & $0.153^{+0.007}_{\mathrm{-0.007}}$ & $3.67^{+0.39}_{\mathrm{-0.38}}$ & 2.6 & 4.7 & 727.2/631\\
[0.15cm]
13 & 300 & $0.80^{+0.21}_{\mathrm{-0.22}}$ & $2.57^{+0.50}_{\mathrm{-0.37}}$ & $1.09^{+0.06}_{\mathrm{-0.06}}$ & $0.152^{+0.006}_{\mathrm{-0.006}}$ & $4.58^{+0.36}_{\mathrm{-0.35}}$ & 0.023 & 3.4 & 570.0/566\\
[0.15cm]
14 & 300 & $1.33^{+0.22}_{\mathrm{-0.23}}$ & $3.67^{+1.21}_{\mathrm{-0.75}}$ & $1.02^{+0.06}_{\mathrm{-0.06}}$ & $0.145^{+0.005}_{\mathrm{-0.005}}$ & $5.20^{+0.43}_{\mathrm{-0.43}}$ & 0.038 & 2.7 & 583.9/528\\
[0.15cm]
15 & 300 & $1.74^{+0.27}_{\mathrm{-0.27}}$ & $5.5^{+4.6}_{\mathrm{-1.7}}$ & $0.95^{+0.06}_{\mathrm{-0.06}}$ & $0.147^{+0.005}_{\mathrm{-0.005}}$ & $5.58^{+0.48}_{\mathrm{-0.51}}$ & 0.060 & 2.3 & 475.1/488\\
[0.15cm]
16 & 277 & $1.49^{+0.31}_{\mathrm{-0.32}}$ & $3.45^{+1.79}_{\mathrm{-0.90}}$ & $0.80^{+0.06}_{\mathrm{-0.06}}$ & $0.144^{+0.004}_{\mathrm{-0.004}}$ & $5.97^{+0.46}_{\mathrm{-0.48}}$ & 0.014 & 1.9 & 519.1/430\\  
\bottomrule
\end{tabular}
\end{table*}

\subsection{GRB 100316D}

GRB 100316D was also discovered by Swift \citep{Stamatikos2010} in the environment of an extended galaxy at the redshift z=0.059 \citep{Vergani2010}. The initial BAT and XRT light curves were very similar to the observed emisson of GRB 060218  \citep{Sakamoto2010} and a thermal component was also observed in X-rays \citep{Starling2012}, although its presence has not been confirmed \citep{Margutti2013}. A SN associated with the burst was also discovered few days after the GRB discovery when it was still rising in luminosity \citep{Chornock2010,Bufano2012}. The similarity between the temporal and spectral properties  of GRB 100316D with those of GRB 060218, makes GRB 100316D an additional test bed for our purposes. Its $T_{\mathrm{90}}$ spectrum, however, is best fitted in the (15 - 150) keV energy range by a simple power-law function with photon index $\gamma = -2.56 \pm 0.18$. We then derive that this GRB is extremely soft, with a peak energy below the lower energy threshold of Swift BAT ($E_{\mathrm{p,i}} \leq 15$ keV).  In analogy with GRB 060218, we have considered the luminous X-ray tail for the computation of the $E_{\mathrm{p,i}}$ and $E_{\mathrm{iso}}$ parameters. However, XRT started to observe GRB 100316D only 144 s after the Swift BAT trigger, and 297 s after the first emission observed by BAT, see Fig. \ref{fig:no2c}.

In order to build an integrated spectrum including both BAT and XRT data, we have simulated the XRT emission in the time interval ($T_0$ -153, $T_0$+144) s, using the \texttt{fakeit} package available in the HEAsoft software packages, and considering the best-fit found for the BAT spectrum. After obtaining an XRT spectrum for the first Swift orbit using the same procedure underlined in the previous section, we have then computed a total integrated spectrum for both detectors by using the \texttt{mathpha} package also available in the HEAsoft suite. The fit of this latter spectrum, with a total exposure time of 891 s, is best fit with an absorbed power-law with an exponential cut-off at $E_{\mathrm{cut}} = 18.7^{+1.1}_{\mathrm{-1.0}}$ keV and a photon index of $\gamma = -1.26^{+0.02}_{\mathrm{-0.02}}$, see also Fig. \ref{tab:no2b}. With these values, we estimate an intrinsic peak energy of $E_{\mathrm{p,i}} = 14.69^{+0.94}_{\mathrm{-0.89}}$ keV and an isotropic energy of $E_{\mathrm{iso}} = 4.841^{+0.026}_{\mathrm{-0.025}} \times 10^{49}$ erg, which implies that GRB 100316D satisfies the Amati relation although its location is borderline (see Fig. 7). Finally, we have built three distinct time-resolved spectra that will be used for the simulation with other detectors. The details of these three time-resolved spectra are shown in Table \ref{tab:no2b}.

\subsection{GRB 161219B}

GRB 161219B has been discovered by Swift BAT \citep{Dai2016} and by Konus WIND \citep{Frederiks2016}. Its redshift has been identified two days later \citep{Tanvir2016} to be $z = 0.1475$ while the emerging supernova was observed 7.24 days after the initial trigger \citep{DeUgarte2016}. The $T_{\mathrm{90}}$ duration observed by Swift BAT is 6.9 s, but a more detailed analysis of BAT data revealed an extended emission, lasting $\sim$ 20 s, anticipating the burst \citep{Palmer2016}, as well as a tail lasting up to 40 s from the GRB trigger, see Fig. \ref{fig:161219B}. Swift XRT started to observe this GRB only 108 s after the BAT trigger \citep{Dai2016}, consequently we do not have a continuity in the observations between BAT and XRT for this GRB.

The $T_{\mathrm{90}}$ spectrum of this GRB, as observed by Swift BAT, is best fitted by a power-law function with an exponential cut off at $E_0 = 92.3^{+68.2}_{\mathrm{-29.0}}$ keV and a photon index of $\gamma = -1.40^{+0.23}_{\mathrm{-0.24}}$  \citep{Cano2017}. The corresponding intrinsic peak energy is $E_{\mathrm{p,i}} = 62.3^{47.0}_{\mathrm{-19.9}}$ keV and the isotropic energy $E_{\mathrm{iso}} = 8.50^{+8.46}_{\mathrm{-3.75}} \times 10^{49}$ erg, in the (1 - 10000) keV energy range. With these values, the location of GRB 161219B is within three sigma of the Amati relation, while if we consider the data provided by the Konus-WIND mission \citep{Frederiks2016}, this burst woould not satisfy the correlation at all, see fig. \ref{fig:Lorenz}.

In order to get more reliable values of the average $E_{\mathrm{p,i}}$ and $E_{\mathrm{iso}}$ of the whole event, we repeated the analysis by including also the first soft/weak pulse and the soft tail described previously and shown in Fig. \ref{fig:161219B}. The BAT data were downloaded, screened and analyzed by following the standard procedures\footnote{The Swift BAT data analysis is described at https://swift.gsfc.nasa.gov/analysis/} and using the usual HEASOFT packages. The total spectrum is best fitted by a Band function \citep{Band1993} with the following parameters: $\alpha = -1.14^{+0.16}_{\mathrm{-0.13}}$, $\beta = -2.37^{+0.52}_{\mathrm{-1.59}}$, and $E_{\mathrm{p,i}} = 55.5^{+14.9}_{\mathrm{-8.9}}$ keV. The total integrated isotropic energy in the (1 - 10000) keV energy range is $E_{\mathrm{iso}} = 1.83 \times 10^{51}$ erg, which puts this GRB well inside the limits of the Amati relation. Finally, we have obtained and analysed four time-resolved spectra from the total emission of GRB 161219B, to be used in the simulations with other detectors. As it is clear from Fig. \ref{fig:161219B}, we have extracted two single spectra from the GRB main pulse and additional two spectra for the precursor and the soft tail. The best fit results of the Swift BAT data for each single spectrum are shown in Table \ref{tab:no2c}.\\

\section{Simulated observations with other detectors}

\begin{figure}
\centering
\includegraphics[scale=0.175]{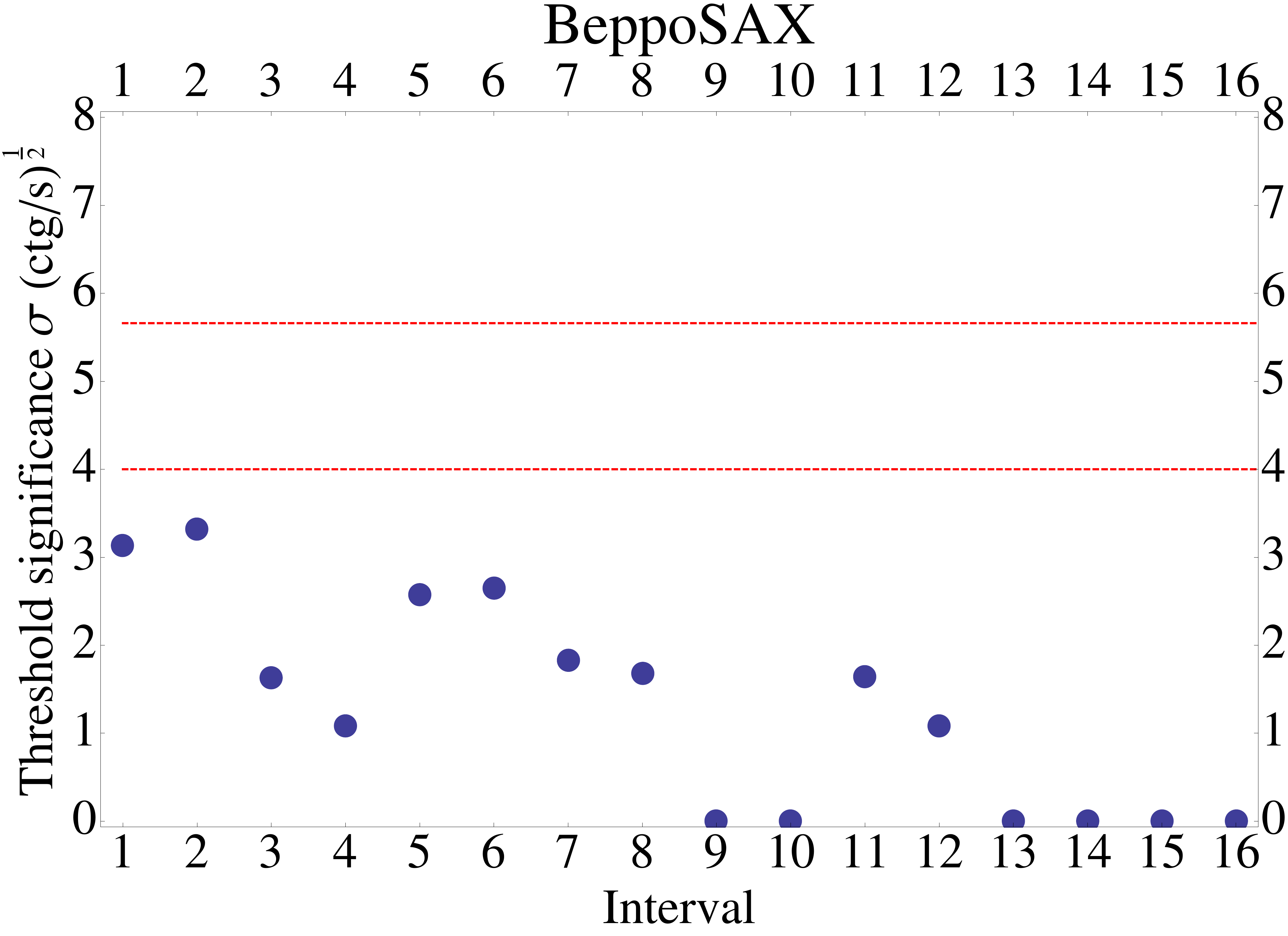}
\includegraphics[scale=0.18]{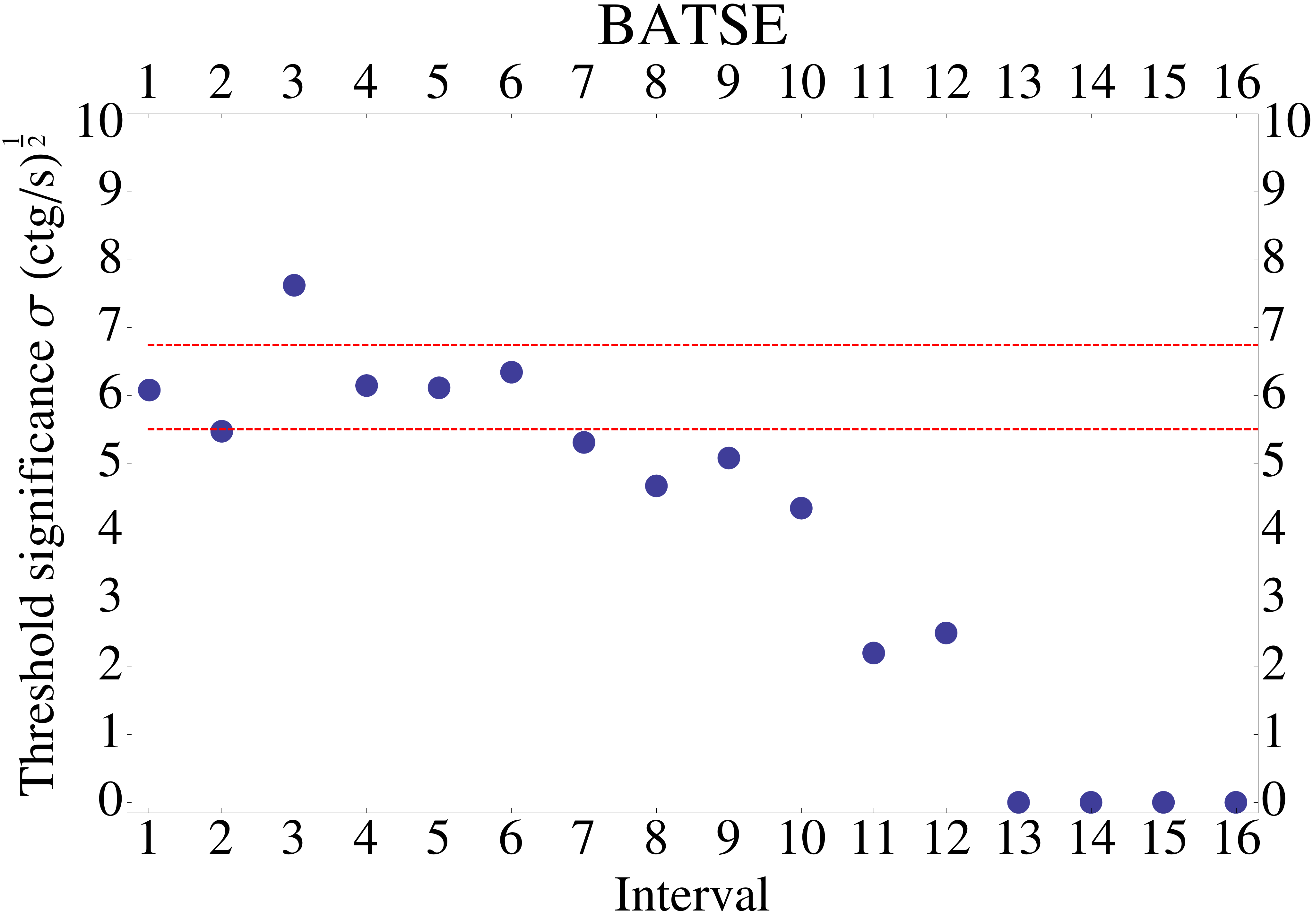}\\
\includegraphics[scale=0.18]{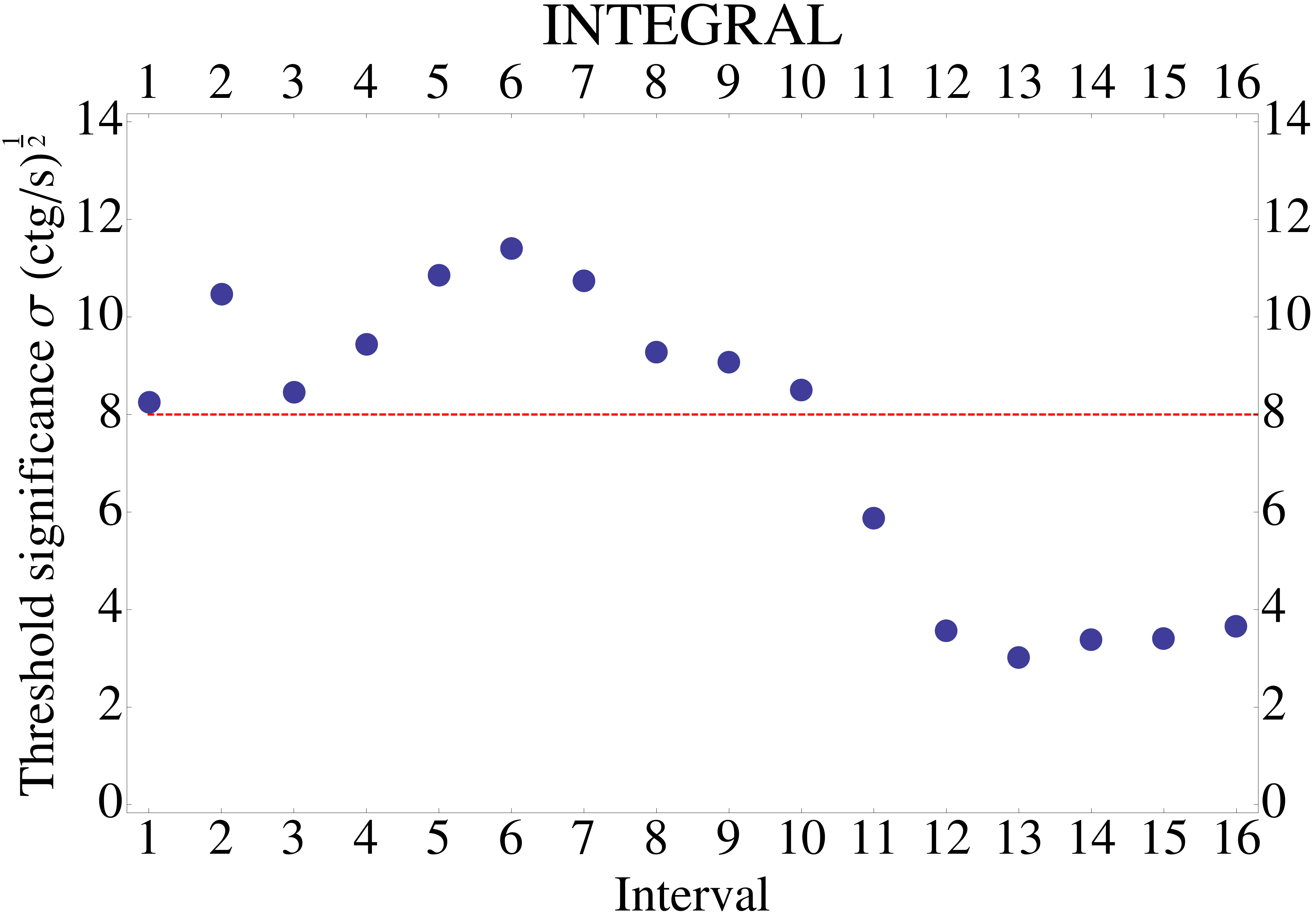}
\includegraphics[scale=0.18]{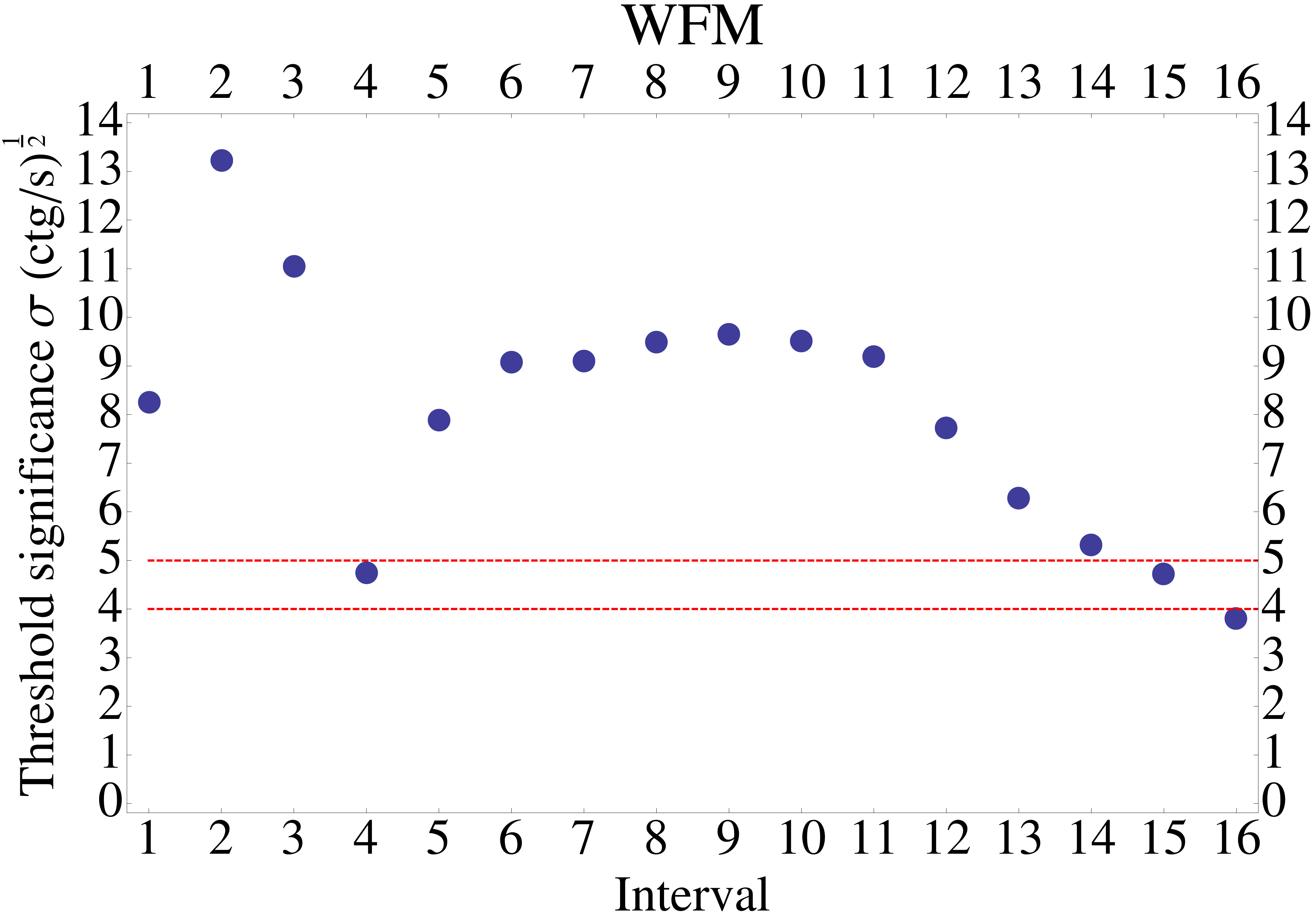}
\caption{Threshold significance $\sigma$ as function of the interval for BeppoSAX, BATSE, INTEGRAL and WFM. The red, horizontal lines represent the $\sigma_{\mathrm{0}}$  threshold as calculated from \ref{eq:no1} (see also \citet{Band1993}. In the case of BeppoSax, BATSE and WFM we report two horizontal lines because the value of the threshold $\sigma_{\mathrm{0}}$ depends on the angle between the direction perpendicular to the plane of the detector and the direction of the source. We use the lower value throughout the whole analysis.}
\label{fig:threshold}
\end{figure}

After deriving the spectral emission of the GRBs 060218, 100316D and 161219B as observed by Swift, we simulated to observe them with old instruments dedicated to GRB observations, as BeppoSAX, BATSE, INTEGRAL, and a planned instrument that is very sensitive to soft X-ray frequencies, as the WFM instrument. 

The energy range of BeppoSAX (see \citet{Frontera2000}) was very wide: from 2 keV to about 700 keV. This large range was obtained thanks to two distinct detectors: the Wide Field Camera (WFC) \citep{Jager1997} which operated between 2 and 30 keV and the Gamma Ray Burst Monitor (GRBM) \citep{Frontera1997}, whose energy range was (40 - 700) keV. In this work, we have considered only the GRBM detector, because it was the GRB alert detector on-board BeppoSAX. The BATSE Large Area Detector (LAD) was an experiment on-board the Compton Gamma-Ray Observer (CGRO), and it consisted of 8 detector module of NaI(TI), covering a wide energy range from 20 keV to 2 MeV. An interesting feature of the BATSE-LAD was that the location of these 8 detectors allowed to cover a very wide fraction of the sky, $\Omega =  4 \pi$. The \emph{INTErnational Gamma-Ray Astrophysics Laboratory} (INTEGRAL) is a facility designed to investigate high-energy objects, carrying detectors for X-ray and gamma-ray part of the spectrum, with energy between 15 keV and 10 MeV \citep{Mereghetti2003}. eXTP \citep{Zhang2016} is a proposed mission for timing analysis of the X-ray transient sky and it should mount also a wide field monitor (WFM) instrument, which is able to detect GRBs in the energy range (2 - 70) keV \citep{LOFT2012}. 

The time-resolved spectral best fits obtained in Sec. 2 (see also Tables \ref{tab:no1}, \ref{tab:no2}, \ref{tab:no2b}, \ref{tab:no2c}) represent our input spectral models in the simulated observations. We used the standard \texttt{fakeit} procedure within the XSPEC package for simulating the observed spectra for all instrument, which requires correct background and spectral response matrix for all detectors, plus an additional ancillary response file for the WFM and INTEGRAL cases. We have obtained the response matrices, background and the ancillary files for each detector from the specific web sites \footnote{http://www.isdc.unige.ch/extp/public-response-files.html}, \footnote{http://saxgrbm.iasfbo.inaf.it/,}, \footnote{http://heasarc.gsfc.nasa.gov/W3Browse/cgro/batsegrbsp.html} or from the literature \citep{Kaneko2006,Guidorzi2011}.

Before spectral fitting, we have grouped any spectra to have a number of ten counts per bin, using the \texttt{grppha} tool of \texttt{heasoft} package. Then, we have used XSPEC to find the best model of each single time-resolved simulated spectrum, as if the GRB was really observed by the considered detector. 

However, in order to obtain the total integrated spectrum and given the different sensitivity of the four detectors considered, we need to consider the effective duration of each GRB emission as observed by each single detector considered. For this reason we have computed the threshold significance $\sigma$ for any single simulated time-resolved spectrum and for each detector considered in order to determine the real duration of the GRB. Following \citet{Band2003}, the threshold significance is given by

\begin{equation}
\label{eq:no1}
\sigma_0 = \frac{A_{eff} f_{det} f_{mask} \Delta t \int_{E_1}^{E_2} \epsilon(E) N(E) d E}{\sqrt{A_{eff} f_{det} \Delta t \int_{E_1}^{E_2} B(E) d E}},
\end{equation}

where $A_{\mathrm{eff}}$ is the effective area of the detector, $f_{\mathrm{det}}$ the fraction of the detector plane that is active, $f_{\mathrm{mask}}$ the fraction of the coded mask that is open, $\Delta t$ the exposure of the photon spectrum $N(E)$, $\epsilon (E)$ the efficiency of the detector and $B(E)$ the background. $E_1$ and $E_2$ correspond respectively to the minimum and the maximum energy threshold for any detector considered in this analysis. 

As final results we obtain different time intervals for each detector considered in which the burst would trigger it, and that also provide a signal with sufficient number of counts to be analyzed with XSPEC, see Fig. \ref{fig:threshold} for the case of GRB 060218: while BeppoSAX would not have triggered at all, WFM would have missed only the last 277 s, and BATSE and INTEGRAL would have seen respectively the first 490 s and 971 s. We have then computed the time-integrated spectra for each detector by summing with \texttt{mathpha} the spectra with a positive detection and then obtained the best-fit for each of them, that resulted to be a cutoff powerlaw in every case. 

Once calculated the $E_{\mathrm{peak}}$ from the simulated spectra, we have estimated the relative $E_{\mathrm{iso}}$ calculating first the corresponding bolometric fluence $S_{\mathrm{bolo}}$ using the relation (see \citealp{Schaefer2007}):

\begin{equation}
\label{eq:Siso}
S_{bolo}=S_{obs}\frac{\int_{\frac{1}{1+z}}^{\frac{10^4}{1+z}}E\phi dE}{\int_{E_{min}}^{E_{max}}E\phi dE},
\end{equation}

being $\Phi$ the differential photon spectrum (dN/dE) and $S_{\mathrm{obs}}$ the observed fluence calculated from the spectrum, $z$ the redshift and $E_{\mathrm{min}}$ and $E_{\mathrm{max}}$ are the extremes of the detector bandpass.
For GRB 060218, we report the result of our calculation in Fig. \ref{fig:Amatiplot_low} and Tables \ref{tab:3b}: while the location estimated with WFM matches with the Amati relation, the ones obtained with BATSE and INTEGRAL do not match it. In the same figure we also report the location of GRB 060218 as observed by Swift BAT+XRT (060218 Swift). On the basis of all these results, we conclude that these different locations in the $E_{\mathrm{p,i}}$-$E_{\mathrm{iso}}$ plane of GRB 060218 as observed by BATSE and INTEGRAL are due to their lack of a highly-sensitive soft X-ray detectors capabilities.  

Similar conclusion are drawn for the cases of GRB 100316D and GRB 161219B. These events show an extended, soft emission similar to that of GRB 060218, although GRB 161219B shows a larger energy output, making us able to explore a different region of the $E_{\mathrm{peak}}/E_{\mathrm{iso}}$ plane in search of the presence a bias effect. The results of our spectral analysis are reported in the Appendix (Tables \ref{tab:no2b} and \ref{tab:no2c}), while the positions on the $E_{\mathrm{peak}}/E_{\mathrm{iso}}$ plane of these two events according to the different detectors we have considered are reported in \ref{fig:Amatiplot_referee}. GRB 161219B results to be an outlier of the Amati relation for all the detectors considered in this work, with the exception of the eXTP-WFM, while for GRB 100316D only BATSE would have measured it outside the Amati relation (but with an unconstrained lower limit for the $E_{\mathrm{p,i}}$ value).

As a countercheck to our result we have applied our approach to a set of eight cosmological bursts (z > 0.1), reported in Table \ref{Longo}, whose $E_{\mathrm{p,i}}$ and $E_{\mathrm{iso}}$  have been measured by Swift-BAT and perfectly match the Amati relation (see Fig \ref{fig:Amatiplot_high}). We performed time-integrated simulations for the BATSE-LAD and BeppoSAX-GRBM instruments and using the observed dataset from Swift. The simulated points in the $E_{\mathrm{p}}$-$E_{\mathrm{iso}}$ plane are reported in Fig \ref{fig:Amatiplot_high}. According to our analysis, in this case we do not see an important effect/bias for these events. All these GRBs are consistent with the $E_{\mathrm{p,i}}$ -- $E_{\mathrm{iso}}$ correlation, even if they would have been observed by BATSE-LAD and BeppoSAX- GRBM. This result implies that this effect is strong only for sub-energetic events and with a soft X-ray prolonged emission.

\begin{table*}
\centering
\caption{Time-integrated spectral fit results for the observed Swift data and for the simulated spectra of GRB 060218}
\label{tab:3b}
\begin{tabular}{l c c c c c c c c c c}     
\hline
$\#$ & $\Delta$t  & $\alpha$ & $E_{\mathrm{p,i}}$ & norm &  $\chi^2$/DOF & $E_{\mathrm{iso}}$ \\
(keV) & ( s ) & ($\gamma$) & ( keV ) &  $\#$ $keV^{-1}$ $cm^{-2}$ $s^{-1}$ & & ($10^{49}$ erg) & \\
\midrule
Swift (BAT+XRT) (0.3-150)  & 2383 & -1.178$_{\mathrm{-0.062}}^{+0.061}$ & 4.62$^{+0.60}_{\mathrm{-0.54}}$ & 0.935$^{+0.017}_{\mathrm{-0.017}}$ & 1083.0/871 & $5.35_{\mathrm{-0.53}}^{+0.53}$  \\
[0.15cm]
Swift (BAT) (15-150) & 2727 & -0.72 $_{\mathrm{-0.12}}^{+0.19} $ & 21.3$^{+3.1}_{\mathrm{-3.5}}$ & 4.8$^{+2.1}_{\mathrm{-2.1}} \times 10^{-3}$ & 1.4/57 & 1.23$^{+0.12}_{\mathrm{-0.11}}$ \\
\midrule
INTEGRAL (15-200)  & 971 & $-1.81_{\mathrm{-0.032}}^{+0.036} $ & $8.4_{\mathrm{-2.0}}^{+2.3}$ & $1.91_{\mathrm{-0.20}}^{+0.20}$ & 0.25/34  & $3.420_{\mathrm{-0.020}}^{+0.023}$ \\
[0.15cm]
BATSE (25-1900) & 490 & $-1.213_{\mathrm{-0.067}}^{+0.091}$ & $33.3_{\mathrm{-7.6}}^{+8.3}$ & $0.032_{\mathrm{-0.095}}^{+0.095}$ & 109.1/112 & $1.268_{\mathrm{-0.024}}^{+0.027}$\\
[0.15cm]
eXTP - WFM (2-50) & 2450 & $-1.831_{\mathrm{-0.012}}^{+0.012}$ & $4.19_{\mathrm{-0.42}}^{+0.46}$ & $1.28_{\mathrm{-0.024}}^{+0.024}$ & 346.8/476 & $4.861_{\mathrm{-0.014}}^{+0.015}$ \\
[0.15cm]
\hline
\end{tabular}
\end{table*}

\subsection{WFM and the class of low-luminosity GRBs-SNe}

\begin{figure}
\centering
\includegraphics[scale=0.25]{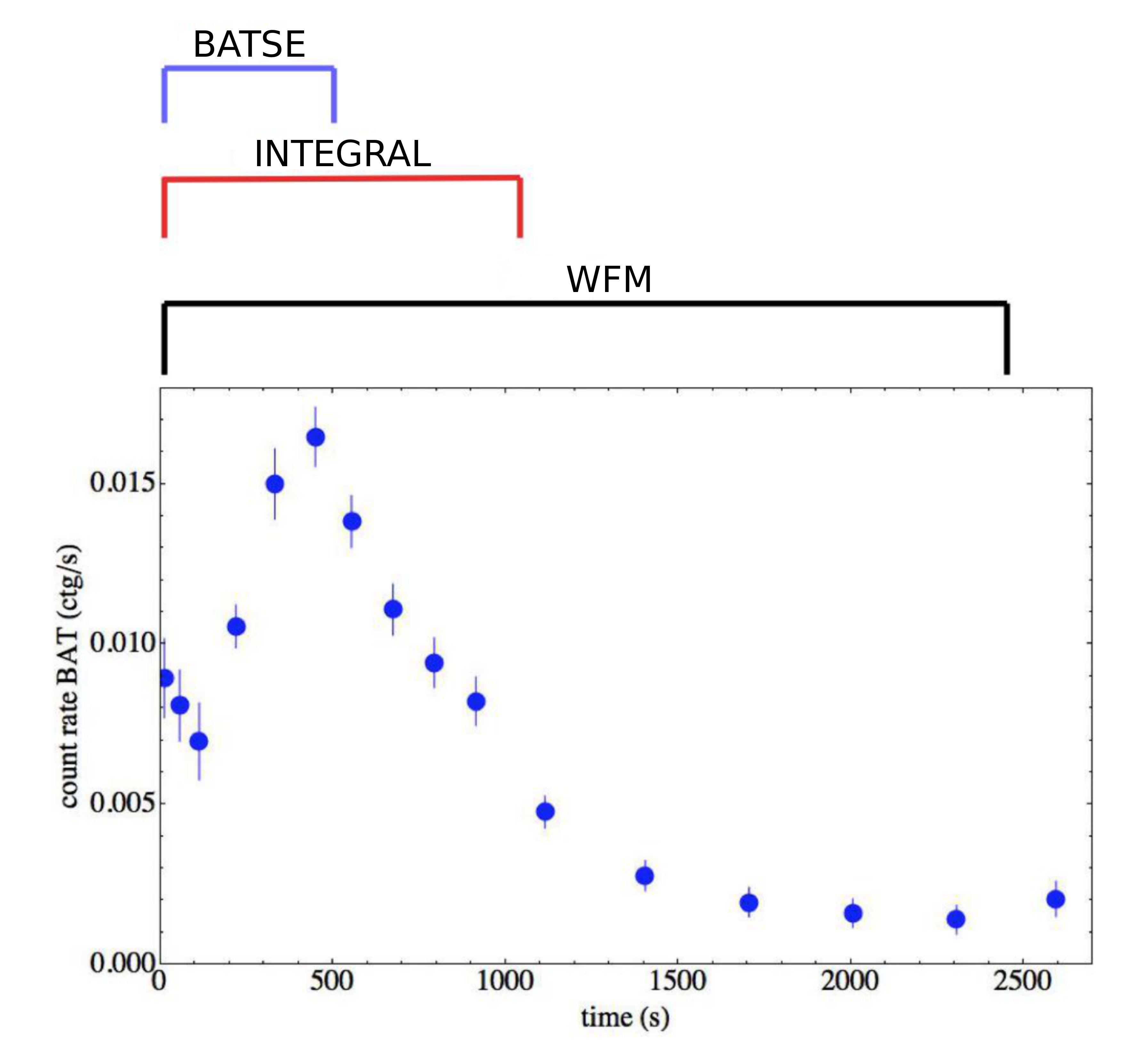}
\caption{The light curve of GRB 060218 as observed by Swift-BAT compared with the effective emission observed by BeppoSAX, BATSE and WFM. }
\label{fig:no3}
\end{figure}

\begin{figure*}
\centering
\includegraphics[scale=0.45]{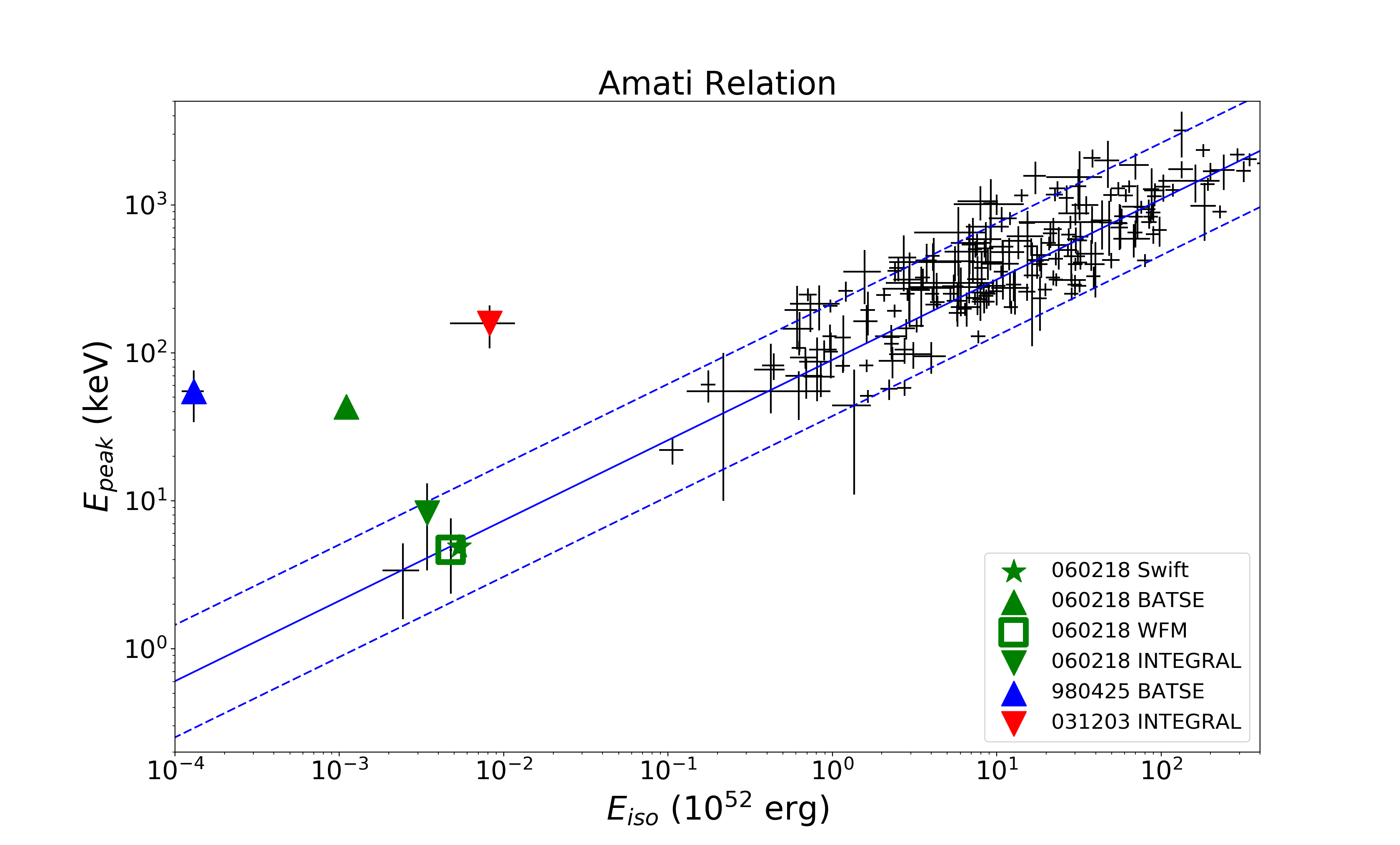}
\caption{The $E_{\mathrm{p,i}}$-$E_{\mathrm{iso}}$ plane (Amati relation). We report in green the position of GRB 060218: according to Swift (BAT+XRT) (star), as it would have been observed by BATSE (triangle), INTEGRAL (reverse triangle) and WFM (square). It is also shown the location in the $E_{\mathrm{p,i}}$-$E_{\mathrm{iso}}$ plane of the two outliers GRB 980425 (blue triangle) and GRB 031203 (red reverse triangle).}
\label{fig:Amatiplot_low}
\end{figure*}

\begin{figure*}
\centering
\includegraphics[scale=0.45]{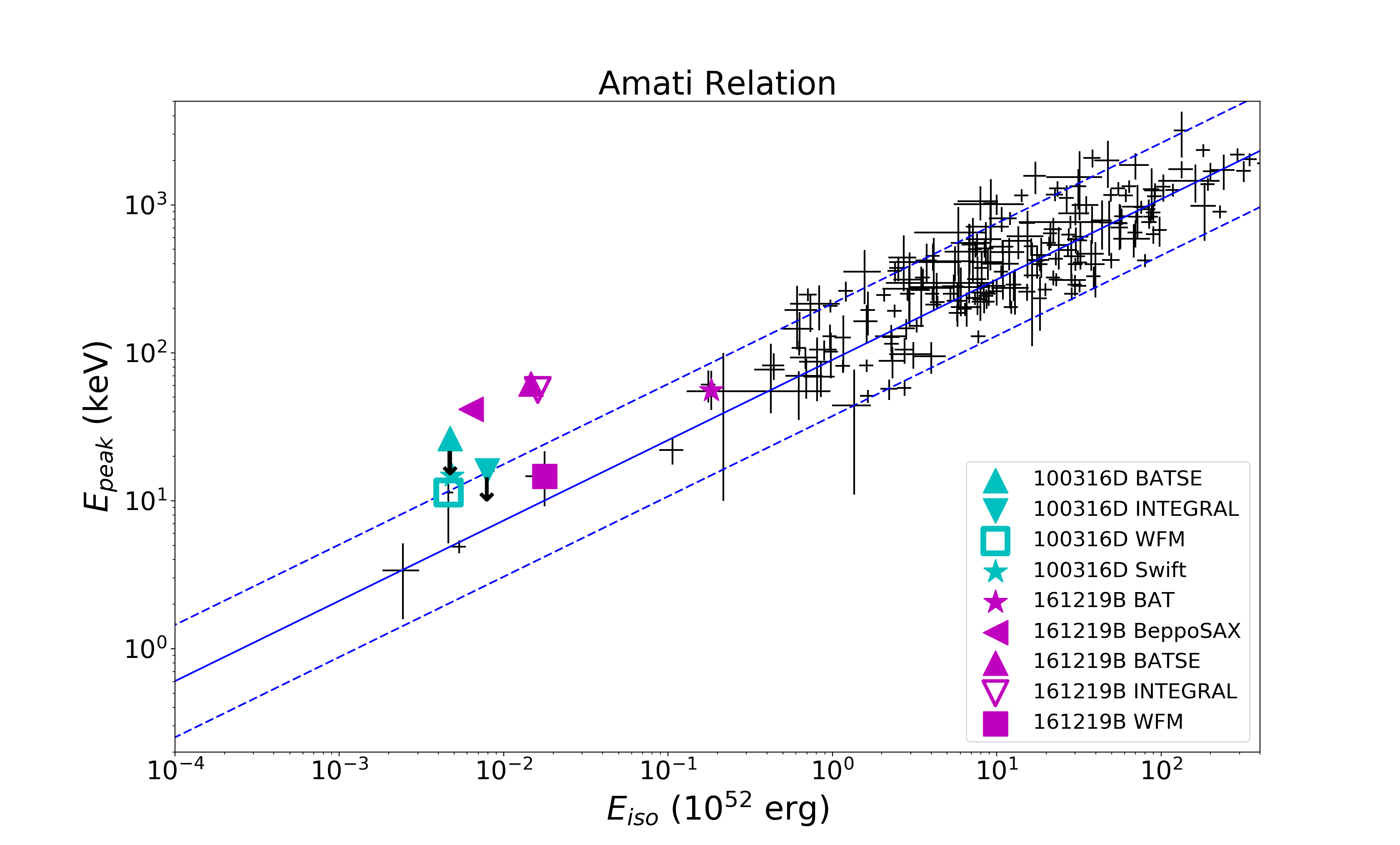}
\caption{The $E_{\mathrm{p,i}}$-$E_{\mathrm{iso}}$ plane (Amati relation). We report in cyan the position of GRB 100316D: according to Swift (BAT+XRT) (star), as it would have been observed by BATSE (triangle - upper limit), INTEGRAL (reverse triangle - upper limit) and WFM (square). It is also shown the location of GRB 161219B: according to Swift (BAT) (star), BeppoSAX (left triangle), BATSE (triangle), INTEGRAL (reverse triangle) and WFM (square - upper limit).}
\label{fig:Amatiplot_referee}
\end{figure*}

We note that during the complete duration of the prompt emission, the value of the peak energy is almost in the energy interval (2 - 70) keV (see fig \ref{fig:no1}), which is the nominal energy range of the Wide Field Monitor proposed for LOFT \citep{LOFT2012} and eXTP \citep{Zhang2016} mission concepts. Its sensitivity, with respect to other current and past GRB detectors, is almost one order of magnitude larger\footnote{http://sci.esa.int/loft/53447-loft-yellow-book/}, suggesting that events similar to GRB 060218 could be detected also at larger distances ($z \approx 0.1-0.2$). It is consequently interesting to estimate the cosmological region of the Universe in which we can detect low-luminosity GRBs-SNe with WFM and, eventually, provide an estimate of the rate of such events. We then have estimated the maximum distance at which GRB 060218-like bursts would still trigger the eXTP-WFM. A positive trigger of eXTP-WFM depends on the assumed  threshold significance, defined in Eq. \ref{eq:no1}, but we need also to correct there the observed spectrum due to the different assumed distance of the GRB. In the specific we modify:
\begin{itemize}
\item the exposure time $\Delta t_z$ varies as $\Delta t_{\mathrm{rest}} (1 + z)$, where $\Delta t_{\mathrm{rest}}$ corresponds to the observed time interval in the rest frame: $\Delta t_{\mathrm{rest}}$ = $\Delta t_{\mathrm{obs}} / 1.0331$;
\item the cutoff energy $E_{\mathrm{cutoff}}$, parameterized as peak energy $E_p$, varies as $E_{\mathrm{cutoff,z}} = E_{\mathrm{cutoff,obs}} \frac{1.0331}{1+z}  $;
\item the normalization of the spectral model varies following the functional form:
\begin{equation}\label{eq:no2}
K_z = K_{\mathrm{obs}} \left( \frac{1+z}{1.0331}\right)^2 \left( \frac{d_l(0.0331)}{d_l(z)} \right)^2.
\end{equation}
\end{itemize}
Note that the subscripts $obs$ correspond to the quantities observed by Swift and the subscript $z$ to the quantities that would be observed if GRB 060218 would stay at redshift $z$. With these corrections we have computed the threshold significance $\sigma$ for the time-resolved spectrum number 6 in Fig. \ref{fig:no3}, which is the spectrum with the largest expected $\sigma$ and not belonging to the initial hard emission of GRB 060218, translated at different redshifts. Assuming a value of $\sigma$ = 4 for the eXTP-WFM, which is the expected final value for the mission, we obtain that an event similar to GRB 060218 would trigger the WFM up to a redshift $z$ = 0.1, see also Fig. \ref{fig:eXTP}.

\begin{table}
\label{Longo}
\caption{Redshift, $E_{\mathrm{peak}}$ and $E_{\mathrm{iso}}$ values obtained from Swift observations.}
\centering
\begin{tabular}{llll}
\toprule
Event & Redshift (z) & $E_{\mathrm{iso}}$ ($10^{52}$ erg) & $E_{\mathrm{peak}}$ (keV) \\
\midrule
GRB 060908 & 1.88 & $7.2_{\mathrm{-1.9}}^{+1.9}$ & $553_{\mathrm{-260}}^{+260}$ \\ [0.15cm]
GRB 060927 & 5.46 & $12.0_{\mathrm{-2.8}}^{+2.8}$ & $275_{\mathrm{-75}}^{+75}$ \\ [0.15cm]
GRB 140206A & 2.74 & $36.806_{\mathrm{-0.058}}^{+0.058}$ & $  364.4_{\mathrm{-4.6}}^{+4.7}$ \\ [0.15cm]
GRB 141220A & 1.34 & $1.6136_{\mathrm{-0.0095}}^{+0.0098}$ & $265_{\mathrm{-11}}^{+12}$ \\ [0.15cm]
GRB 151029A & 2.74 & $8.013_{\mathrm{-0.068}}^{+0.069}$ & $117.6_{\mathrm{-4.1}}^{+4.1}$ \\ [0.15cm]
GRB 160227A & 2.38 & $5.924_{\mathrm{-0.038}}^{+0.039}$ & $248_{\mathrm{-14}}^{+15}$ \\ [0.15cm]
GRB 161117A & 1.55 & $14.858_{\mathrm{-0.018}}^{+0.018}$ & $145.2_{\mathrm{-1.2}}^{+1.2}$ \\ [0.15cm]
GRB 170113A & 1.97 & $0.7299_{\mathrm{-0.0052}}^{+0.0053}$ & $200.9_{\mathrm{-8.8}}^{+9.1}$ \\ [0.15cm]
\bottomrule
\end{tabular}
\end{table}


We have also determined at which redshift GRB 060218 would have been observed by Swift BAT and we have obtained a redshift of $z = 0.05$ as the detection limit for GRB 060218 with Swift BAT.
The cosmological comoving element volume at redshift $z$ is given by
\begin{equation}
V(z) = \frac{4 \pi}{3}  d_c(z)^3,
\end{equation}
where $d_c = d_l /(1+z)$ the cosmological comoving distance.
At these distances ($z=0.1$), the comoving element volume is 30 times larger than the one at $z=0.0331$ and eight times larger than the one at $z = 0.05$.

\section{Conclusions}

\begin{figure*}
\centering
\includegraphics[scale=0.45]{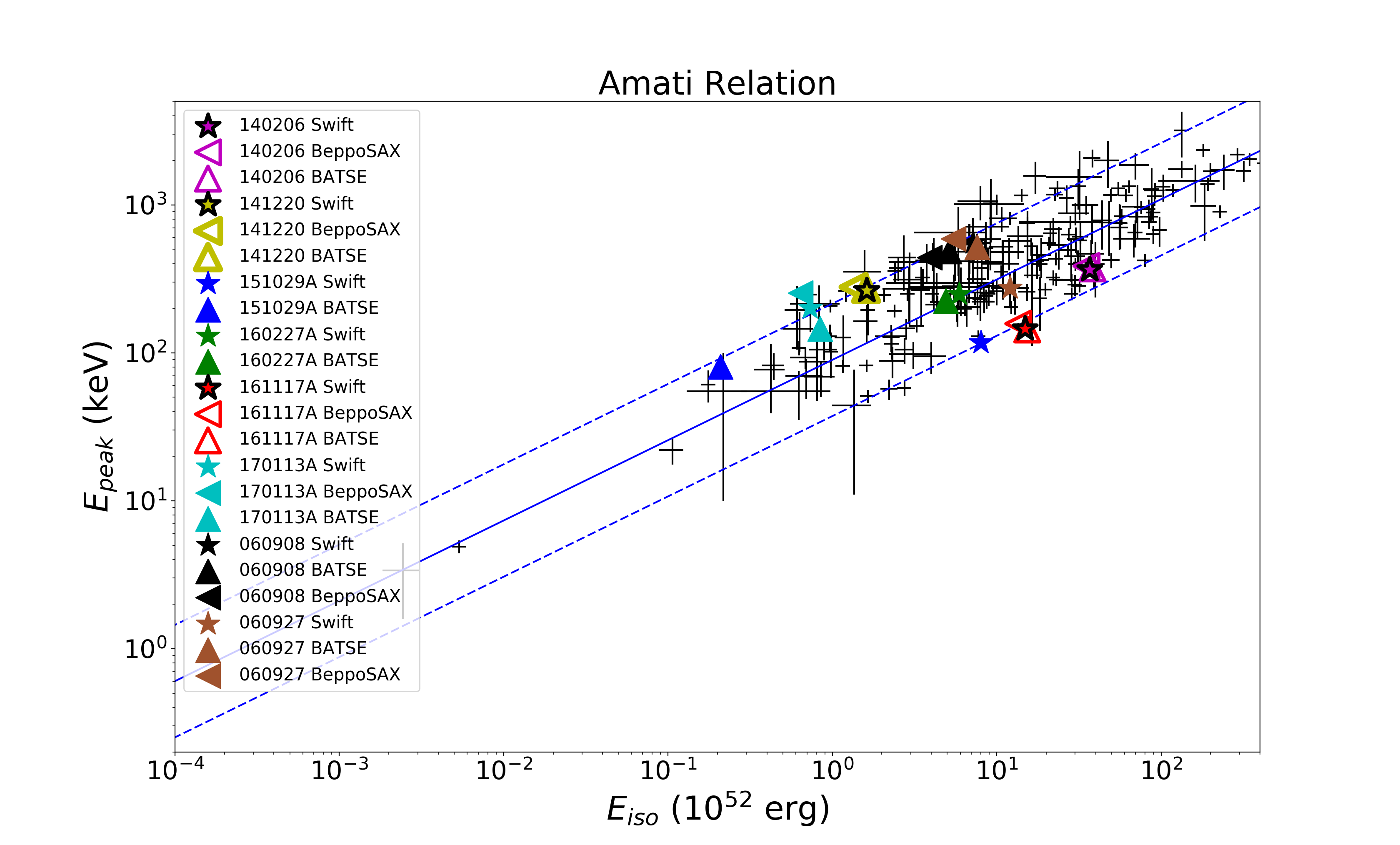}
\caption{The positions in the Ep,i-Eiso plane of the eight cosmological GRBs, described in the text, characterized by a higher energetic and a larger redshift than GRB 060218, GRB 161219B and GRB 100316D. We note how the emission of these events, as observed by BATSE and Beppo-SAX will still satisfy the $E_{\mathrm{p,i}}$-$E_{\mathrm{iso}}$ correlation.}
\label{fig:Amatiplot_high}
\end{figure*}

The main results of our analysis can be summarized as follows: 

i) GRB 060218, GRB 161219B and 100316D if observed with detectors not sensitive at low energies ( $\sim$ 0.3 keV) such as INTEGRAL and/or  BATSE, would be outliers of the Amati relation \citealp[see][for a quantitative analysis of the instumental bias]{AmatiDichiara13}. On the other hand GRB 060218 and GRB 100316D perfectly match the $E_{\mathrm{p,i}} - E_{\mathrm{iso}}$ relation after being observed with Swift (down to 0.3 keV). On the basis of this result we suggest that GRB 980425 and GRB 031203 are not ``true'' outliers of the Amati relation and their location in the $E_{\mathrm{p,i}} - E_{\mathrm{iso}}$ plane is the result of an observational bias, rather than being related to a combination of the geometry of GRB explosions with the line of sight of the observer \citep[e.g. GRB viewed off-axis][]{Yamazaki2004,RamirezRuiz2005,Eichler2004};

ii) the above conclusion is strengthened by the fact that Swift-BAT (15-150 keV) actually measured GRB 060218 as an outlier and XMM-Newton observed an X-ray echo that suggests the presence of an extended-soft emission associated to GRB 031203 \citep{Watson20062};

iii) in the case of GRB 100316D, we note that the WFM observes it at the border of the 1-$\sigma$ boundary of the Amati relation. We consequently derive that it is not sufficient to observe below to the limit of the soft X-rays energy range ($\sim 0.3$ keV), but it is necessary to use a detector with large sensitivities at these energies, like Swift-XRT, in order to get as much information as possible about the total energetic emitted by these low-luminosity GRBs;

iv) to give more weight to our conclusions, we have applied the same approach to a sample of "high-luminosity" GRBs whose $E_{\mathrm{p,i}}$ and $E_{\mathrm{iso}}$ parameters, reported in Table \ref{Longo}, have been measured by Swift-BAT. All these GRBs are more energetic and located at higher redshift than GRB 060218, GRB 100316D and GRB 161219B, and then are not expected to show the X-ray soft tail. In these cases the GRBs always match the $E_{\mathrm{p,i}} - E_{\mathrm{iso}}$ relation either if observed by Swift or BATSE or BeppoSAX (see Fig. \ref{fig:Amatiplot_high});

v) after simulating WFM observations, we have shown that GRB 060218 could have been observed up to $z = 0.1$ which is about three times farther than it was observed with Swift-BAT \citep{Guetta2007}. As a consequence, we are likely missing a significant fraction of low-luminosity and sub-energetic GRBs, whose high-energy emission remains undetected due to the poor sensitivity and limits of current operating detectors. 

\begin{table*}
\centering
\caption{Time-integrated spectral fit results for the observed Swift data and for the simulated spectra of GRB 100316D}
\label{tab:3b}
\begin{tabular}{l c c c c c c c c c c}     
\hline
$\#$ & $\Delta$t  & $\alpha$ & $E_{\mathrm{p,i}}$ & norm &  $\chi^2$/DOF & $E_{\mathrm{iso}}$ \\
(keV) & ( s ) & ($\gamma$) & ( keV ) &  $\#$ $keV^{-1}$ $cm^{-2}$ $s^{-1}$ & & ($10^{49}$ erg) & \\
\midrule
Swift (BAT+XRT) (0.3-150) & 890 & -1.256 $_{\mathrm{-0.018}}^{+0.018} $ & 14.69$^{+0.94}_{\mathrm{-0.89}}$ & 0.4446 $^{+0.0053}_{\mathrm{-0.0053}} $ & 568.7/971 & 4.841$^{+0.026}_{\mathrm{-0.025}}$ \\
\midrule
INTEGRAL (15-200)  & 497 & $-2.38_{\mathrm{-0.23}}^{+0.26} $ & $15.9$ & $4.9_{\mathrm{-2.6}}^{+6.1}$ & 1.8/35  & $7.91_{\mathrm{-0.51}}^{+0.92}$ \\
[0.15cm]
BATSE (25-1900) & 200 & $-2.358_{\mathrm{-0.047}}^{+0.059}$ & $26.5$ & $6.9_{\mathrm{-1.5}}^{+1.5}$ & 102/113 & $4.722_{\mathrm{-0.097}}^{+0.104}$\\
[0.15cm]
eXTP - WFM (2-50) & 891 & $-1.62_{\mathrm{-0.14}}^{+0.14}$ & $11.4_{\mathrm{-6.2}}^{+17.7}$ & $0.710_{\mathrm{-0.041}}^{+0.084}$ & 415/476 & $4.61_{\mathrm{-0.14}}^{+0.33}$ \\
[0.15cm]
\hline
\end{tabular}
\end{table*}

\begin{table*}
\centering
\caption{Time-integrated spectral fit results for the observed Swift data and for the simulated spectra of GRB 161219B}
\label{tab:3b}
\begin{tabular}{l c c c c c c c c c c}     
\hline
$\#$ & $\Delta$t  & $\alpha$ & $E_{\mathrm{p,i}}$ & norm &  $\chi^2$/DOF & $E_{\mathrm{iso}}$ \\
(keV) & ( s ) & ($\gamma$) & ( keV ) &  $\#$ $keV^{-1}$ $cm^{-2}$ $s^{-1}$ & & ($10^{49}$ erg) & \\
\midrule
Swift (BAT) (15-150) & 62 & -1.435 $_{\mathrm{-0.042}}^{+0.050} $ & 56 $^{+20}_{\mathrm{-14.0}}$ & 20.1$^{+3.4}_{\mathrm{-3.4}} $ & 1.2/55 & 183 $^{+5.1}_{\mathrm{-4.2}}$ \\
\midrule
BeppoSAX (40-700)  & 8 & $-1.46_{\mathrm{-0.72}}^{+0.60} $ & $41.5_{\mathrm{-1.9}}^{+1.9}$ & $13.44 _{\mathrm{-0.76}}^{+0.76}$ & 339/223  & $6.328_{\mathrm{-0.038}}^{+0.039}$ \\
[0.15cm]
INTEGRAL (15-200)  & 62 & $-1.5304_{\mathrm{-0.0058}}^{+0.0059} $ & $56.0_{\mathrm{-3.2}}^{+3.4}$ & $2.1506_{\mathrm{-0.0058}}^{+0.0059}$ & 5.9/34  & $16.127_{\mathrm{-0.057}}^{+0.060}$ \\
[0.15cm]
BATSE (25-1900) & 62 & $-1.336_{\mathrm{-0.012}}^{+0.013}$ & $61.8_{\mathrm{-3.6}}^{+3.8}$ & $1.202_{\mathrm{-0.063}}^{+0.063}$ & 38.15/112 & $1.465_{\mathrm{-0.097}}^{+0.100}$\\
[0.15cm]
eXTP - WFM (2-50) & 62 & $-1.7297_{\mathrm{-0.0051}}^{+0.0053}$ & $14.7_{\mathrm{-4.2}}^{+3.6}$ & $1.173_{\mathrm{-0.023}}^{+0.023}$ & 475/478 & $17.72_{\mathrm{-4.15}}^{+0.36}$ \\
[0.15cm]
\hline
\end{tabular}
\end{table*}

\begin{acknowledgements}
This research has made use of data, software and web tools obtained from the High Energy Astrophysics Science Archive Research Center (HEASARC), a service of the Astrophysics Science Division at NASA/GSFC and of the Smithsonian Astrophysical Observatory's High Energy Astrophysics Division.
This work made use of data supplied by the UK Swift Science Data Centre at the University of Leicester.
We thank the anonymous referee for his useful comments and suggestions that improved our paper.
We thank Sergio Campana for providing us the Swift BAT spectral dataset of GRB 060218.
A special thank goes to Prof. Mauro Orlandini that contributed to the statistical analysis contained in the work and Cristiano Guidorzi for his useful suggestions.
LI7 acknowledges support from the Spanish research project AYA 2014-58381-P.
\end{acknowledgements}

 \bibliographystyle{aa}

\begin{appendix}
\section{Tables and Figures}
In this section we report the lightcurves and the statistical results of spectral analysis performed on GRB 100316D and GRB 161219B. We use these two events as a further verification of our thesis on the behaviour of the statistical bias we describe in the conclusions of our work.

\begin{table*}
\centering
\tiny
\caption{Swift BAT+XRT (0.3-150 keV) spectral fits data results of the GRB 100316D dataset. For the first spectrum, we used BAT data only.}
\label{tab:no2b}
\begin{tabular}{l c c c c c c c c c}     
\toprule
$\#$ & $\Delta$t   & $\gamma$ & $E_{\mathrm{cutoff}}$  & norm CPO & flux BAT & flux XRT & $\chi^2$/DOF\\
& & & & & (10$^{-9}$) & (10$^{-9}$) &\\
& (s)	 & & (keV)  & Photons $keV^{-1}$ $cm^{-2}$ $s^{-1}$ & (erg/cm$^2$/s) & (erg/cm$^2$/s) &   \\
\midrule
1 & 297 & $1.75^{+0.84}_{\mathrm{-0.68}}$ & $45.3^{+48.5}_{\mathrm{-21.2}}$ & $1.31^{+2.77}_{\mathrm{-1.30}}$ & 4.13 & -- & 63.7/55\\
[0.15cm]
2 & 200 & $1.32^{+0.03}_{\mathrm{-0.03}}$ & $35.7^{+5.18}_{\mathrm{-4.35}}$ & $0.28^{+0.01}_{\mathrm{-0.01}}$ & 3.97 & 2.23 & 856.2/814\\
[0.15cm]
3 & 394 & $1.29^{+0.03}_{\mathrm{-0.03}}$ & $20.7^{+2.59}_{\mathrm{-2.28}}$ & $0.28^{+0.01}_{\mathrm{-0.01}}$ & 2.07 & 2.04 & 880.6/895\\
\bottomrule
\end{tabular}
\end{table*}

\begin{figure}
\centering
\includegraphics[scale=0.32]{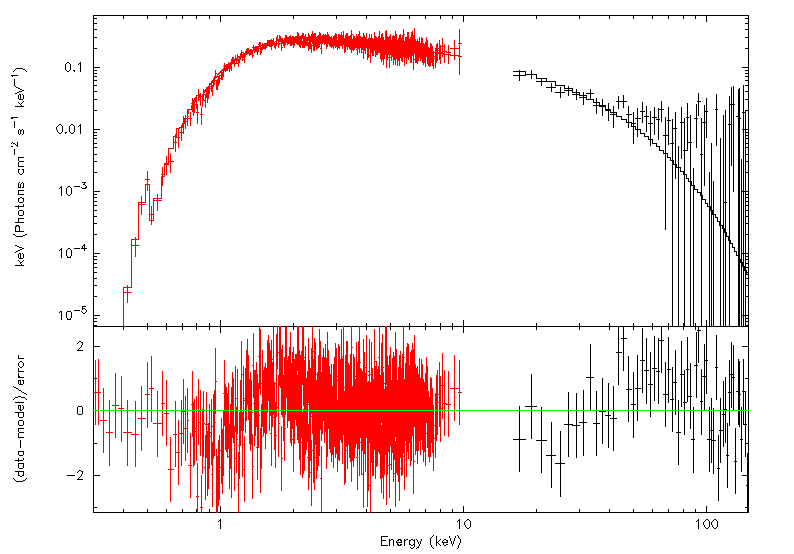}
\caption{Best-fit of the Swift BAT+XRT integrated spectrum of GRB 100316D with an absorbed power-law function with an exponential cut off.}
\label{fig:no2b}
\end{figure} 

\begin{figure}
\centering
\includegraphics[scale=0.37]{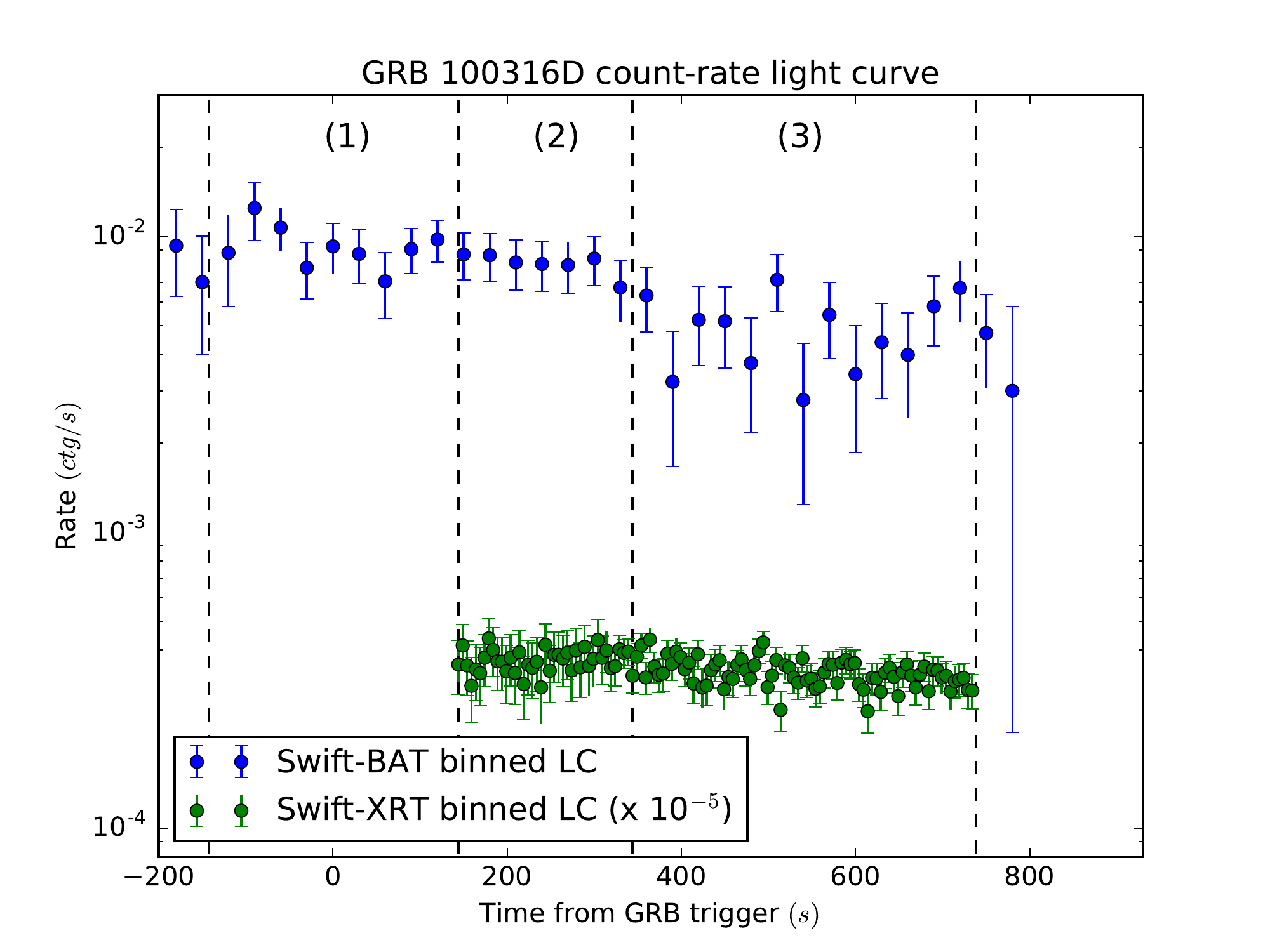}
\caption{Swift BAT (blue circles) and XRT (green circles) count-rate light curve of GRB 100316D. Both curves have been binned while the XRT one has also been rescaled by $10^{-5}$. The dashed black lines mark the time intervals of the time-resolved spectra considered in our analysis (see also Table \ref{tab:no2b})}
\label{fig:no2c}
\end{figure} 

\begin{table*}
\centering
\tiny
\caption{Swift BAT (15-150 keV) spectral fits data results of the GRB 161219B dataset.}
\label{tab:no2c}
\begin{tabular}{l c c c c c c c c c}     
\toprule
$\#$ & $\Delta$t   & $\gamma$ & $E_{\mathrm{cutoff}}$  & norm CPO & flux BAT & $\chi^2$/DOF\\
& & & & & (10$^{-9}$) & (10$^{-9}$) &\\
& (s)	 & & (keV)  & Photons $keV^{-1}$ $cm^{-2}$ $s^{-1}$ & (erg/cm$^2$/s) &   \\
\midrule
1 & 21 & $2.20^{+1.55}_{\mathrm{-0.85}}$ & -- & $3.0^{+8.3}_{\mathrm{-2.9}}$ & 5.1 & 51.6/56\\
[0.15cm]
2 & 3 & $1.51^{+0.28}_{\mathrm{-0.31}}$ & $112.8^{+212.5}_{\mathrm{-47.9}}$ & $11.5^{+13.5}_{\mathrm{-6.6}}$ & 170.8  & 79.8/77\\
[0.15cm]
3 & 4 & $1.21^{+0.29}_{\mathrm{-0.32}}$ & $68.0^{+51.1}_{\mathrm{-22.2}}$ & $4.7^{+5.8}_{\mathrm{-2.7}}$ & 156.7  & 84.9/77\\
[0.15cm]
4 & 33 & $2.05^{+0.35}_{\mathrm{-0.32}}$ & --& $2.8^{+6.3}_{\mathrm{-2.7}}$ & 8.4  & 55.2/78\\
\bottomrule
\end{tabular}
\end{table*}

\begin{figure}
\centering
\includegraphics[scale=0.37]{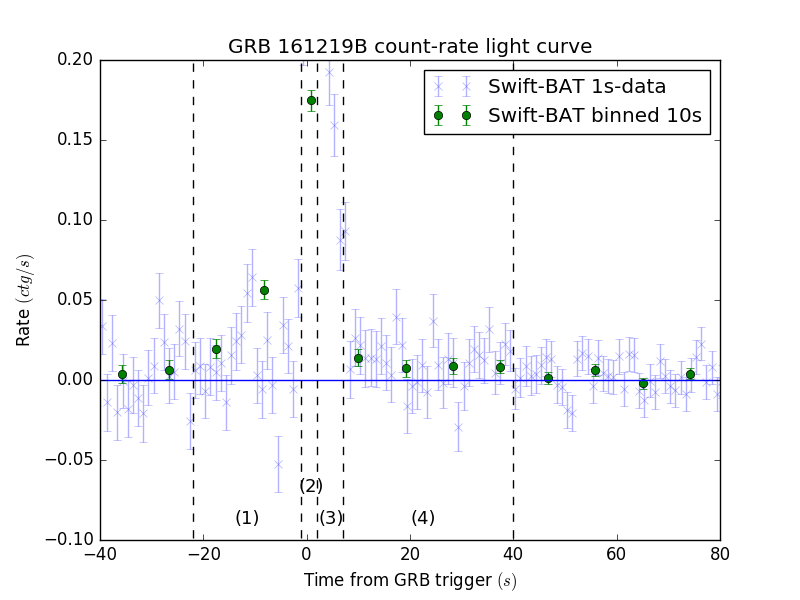}
\caption{Swift BAT count-rate light curve of GRB 100316D binned at 1 s (blue data) and at 10 s (green circles). The dashed black lines mark the time intervals of the time-resolved spectra considered in our analysis (see also Table \ref{tab:no2c})}
\label{fig:161219B}
\end{figure} 

\begin{figure}
\centering
\includegraphics[scale=0.30]{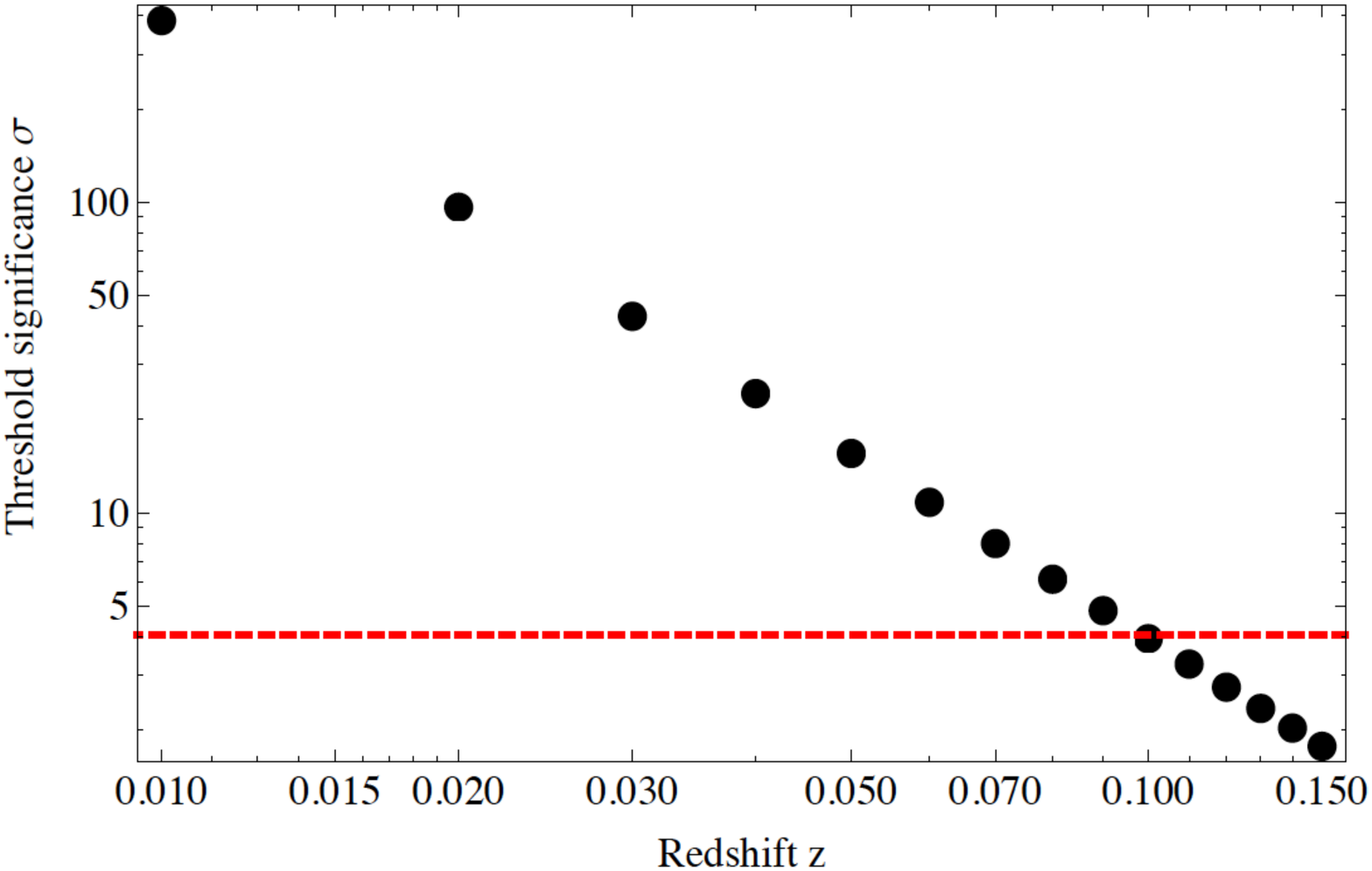}
\caption{The eXTP threshold significance for the most bright time-resolved spectrum of GRB 060218 (number 6) as a function of the redshift. The eXTP detector would trigger on GRB 060218 up to redshift $z = 0.01$.}
\label{fig:eXTP}
\end{figure}

\section{Simulated Lightcurves}

In this section we report the simulated lightcurves for the three instrument that we analyze and that could have led to a positive detection of an emission like that of GRB 060218. These instruments could have detected only a fraction of the total emission. The x axis has covers the total time extension of the dataset we have analyzed.

\begin{figure}
\centering
\includegraphics[scale=0.22]{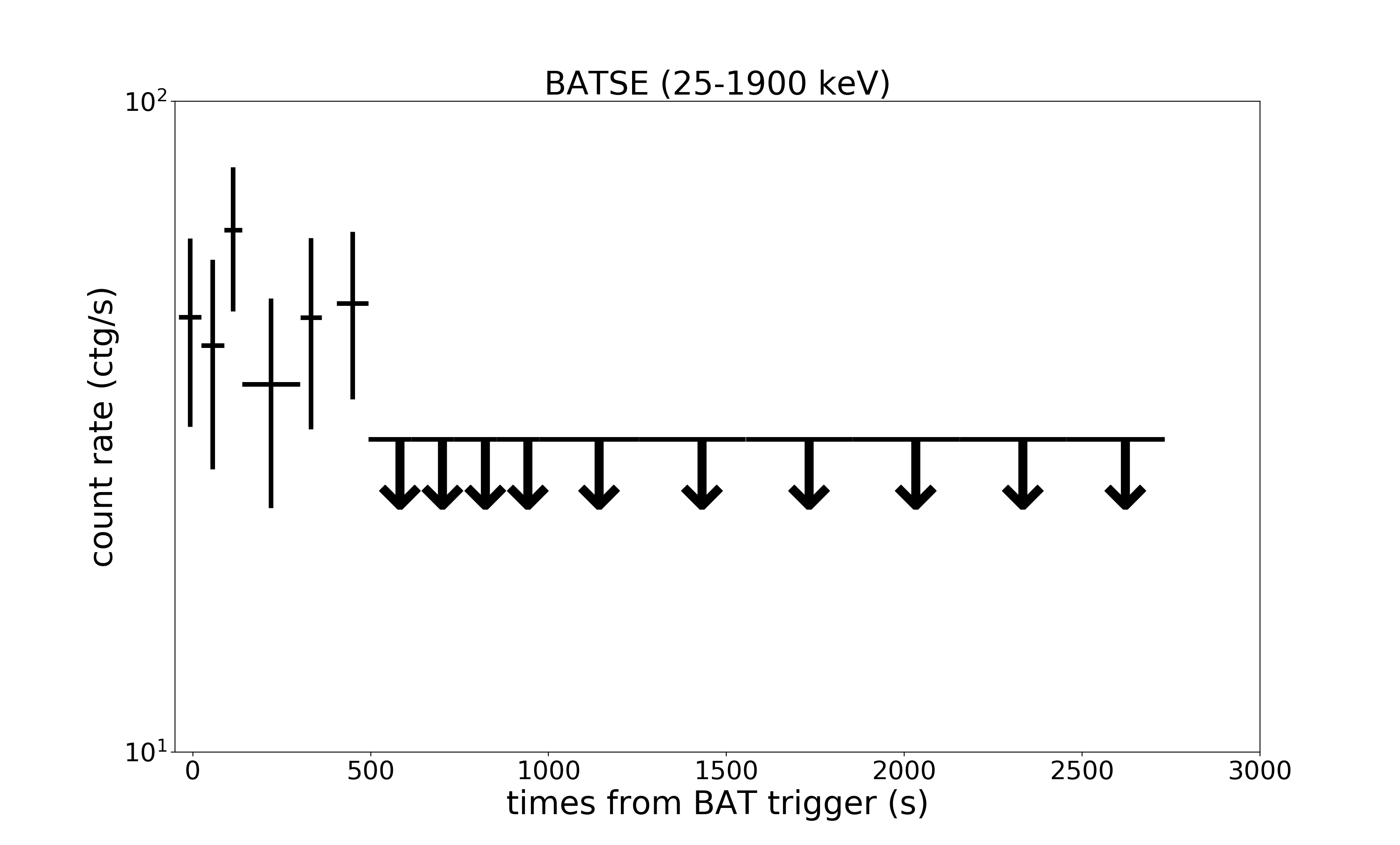}\\
\includegraphics[scale=0.22]{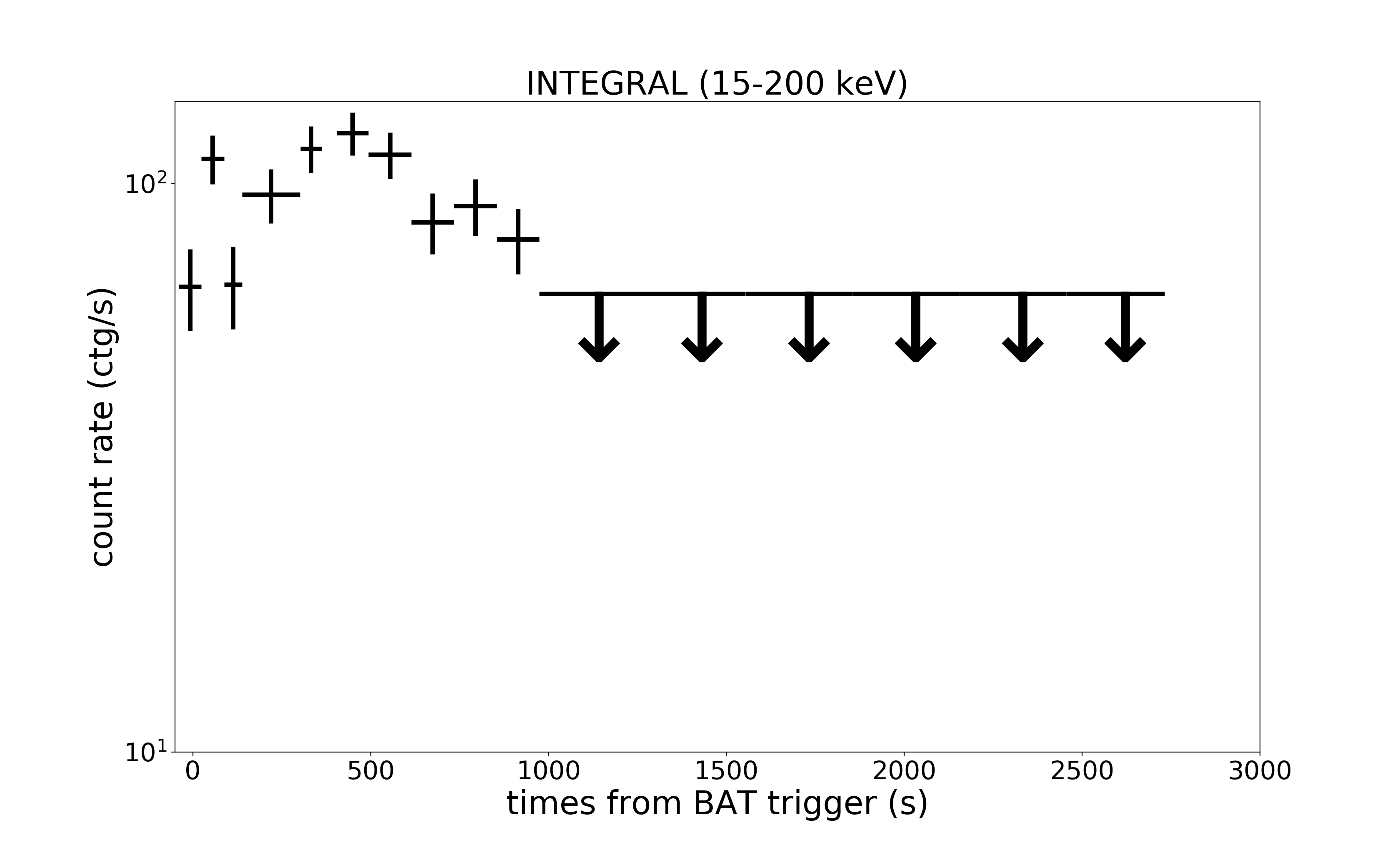}\\
\includegraphics[scale=0.22]{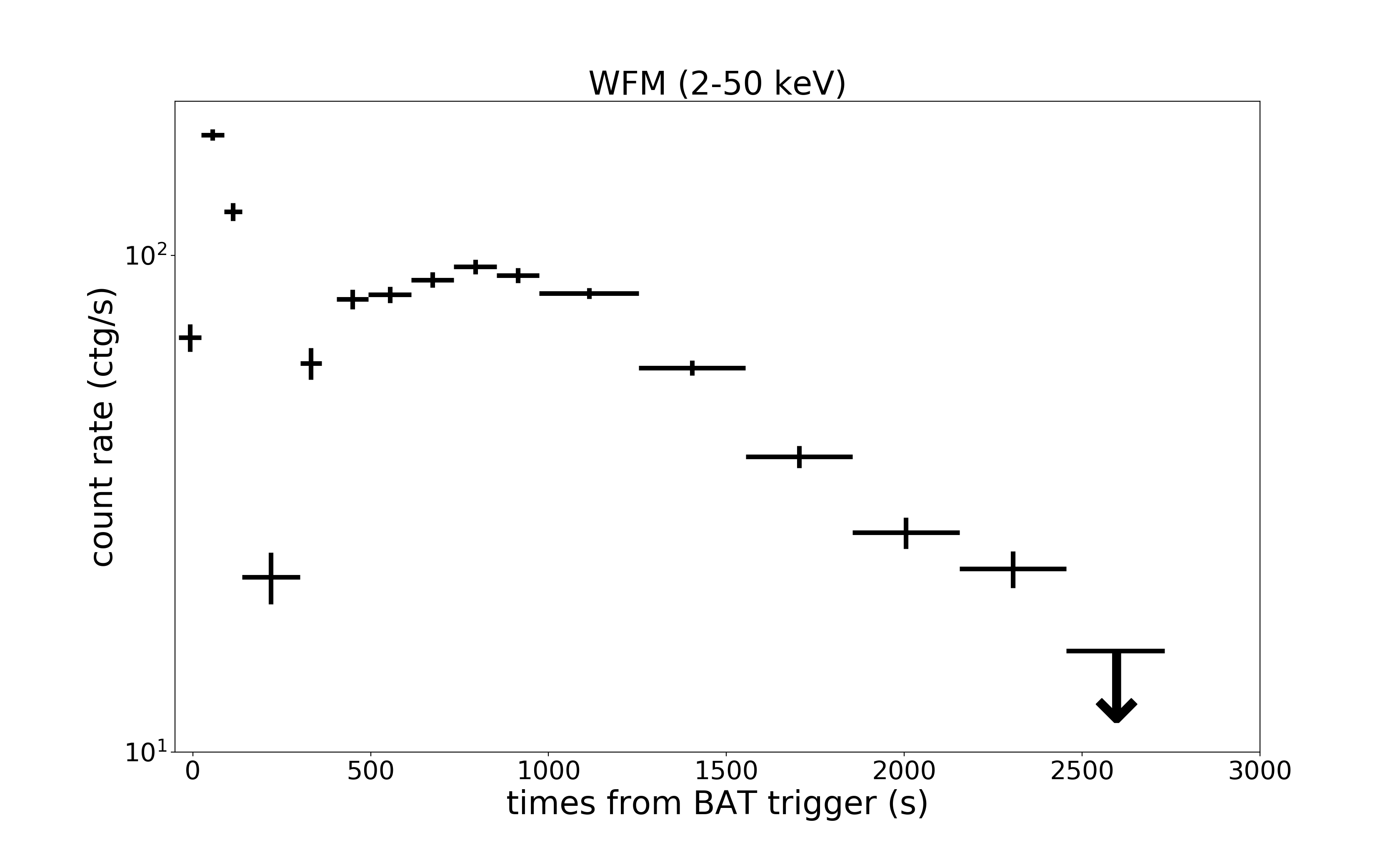}
\caption{Lightcurve of GRB 060218 as seen by BATSE, INTEGRAL and the WFM-eXTP according to our simulations. The last intervals of our analysis would have been under the detection threshold of the instrument and so no counts are expected.}
\label{fig:LightCurve_INTEGRAL}
\end{figure} 

\begin{figure}
\centering
\includegraphics[scale=0.22]{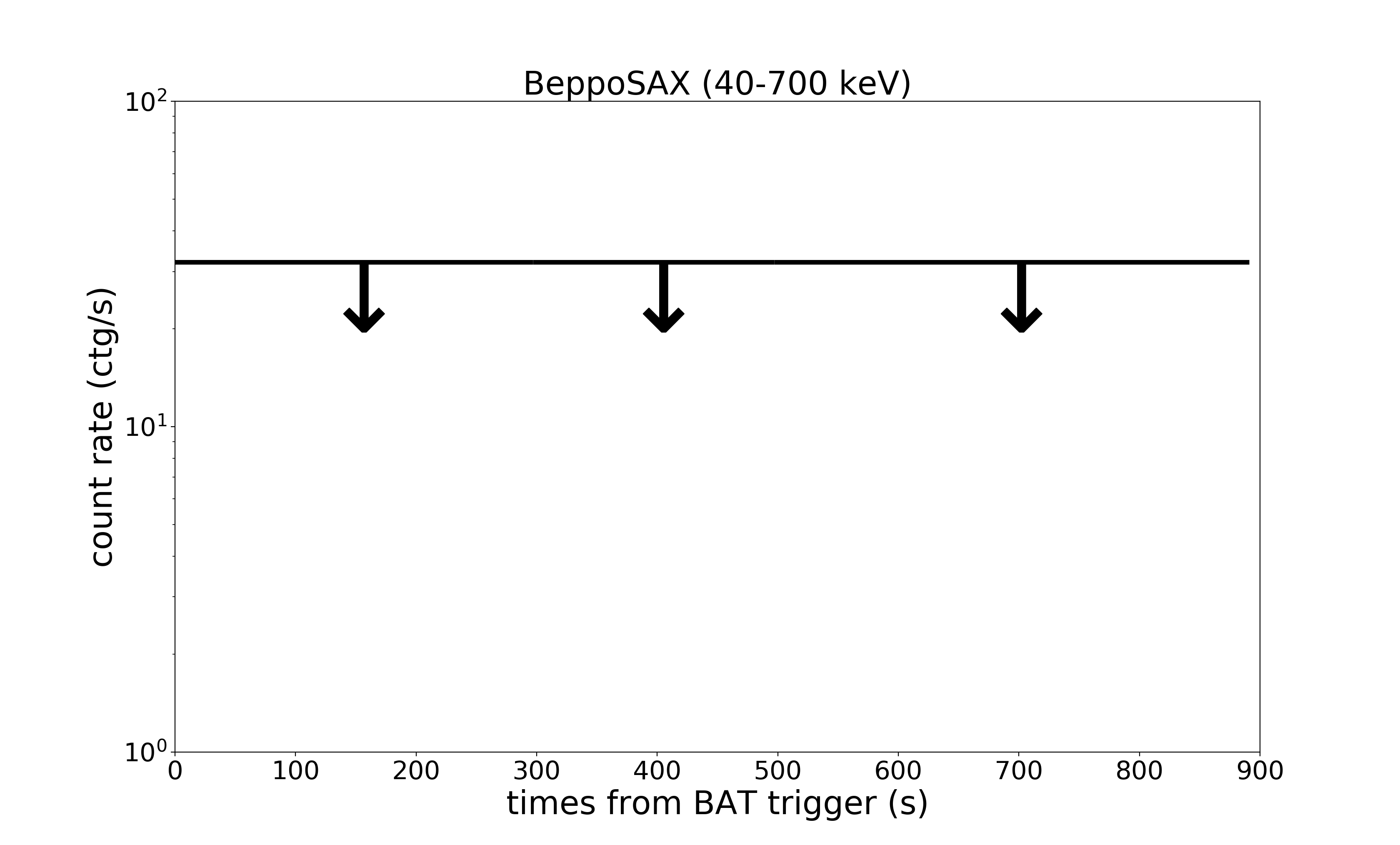}\\
\includegraphics[scale=0.22]{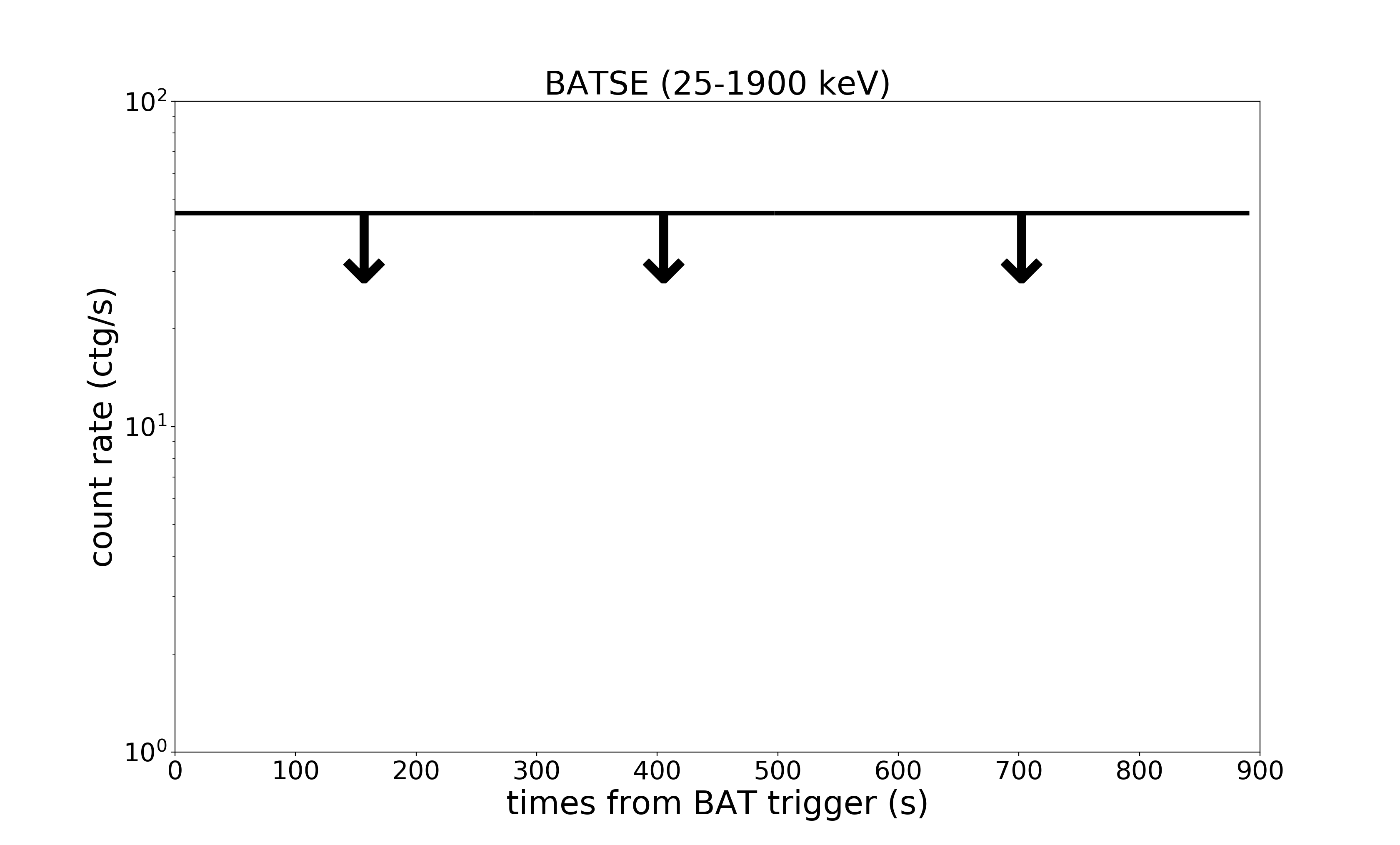}\\
\includegraphics[scale=0.22]{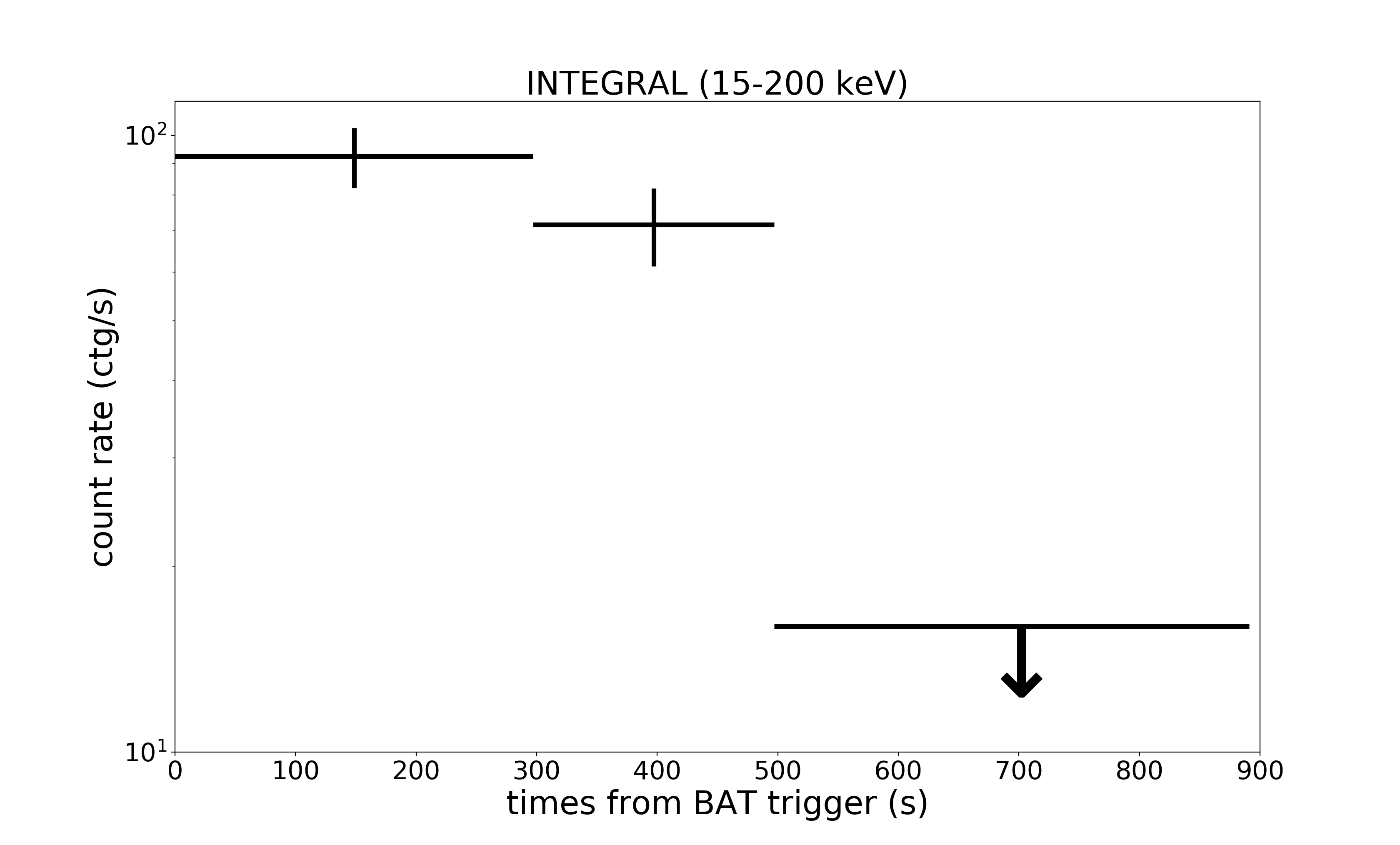}\\
\includegraphics[scale=0.22]{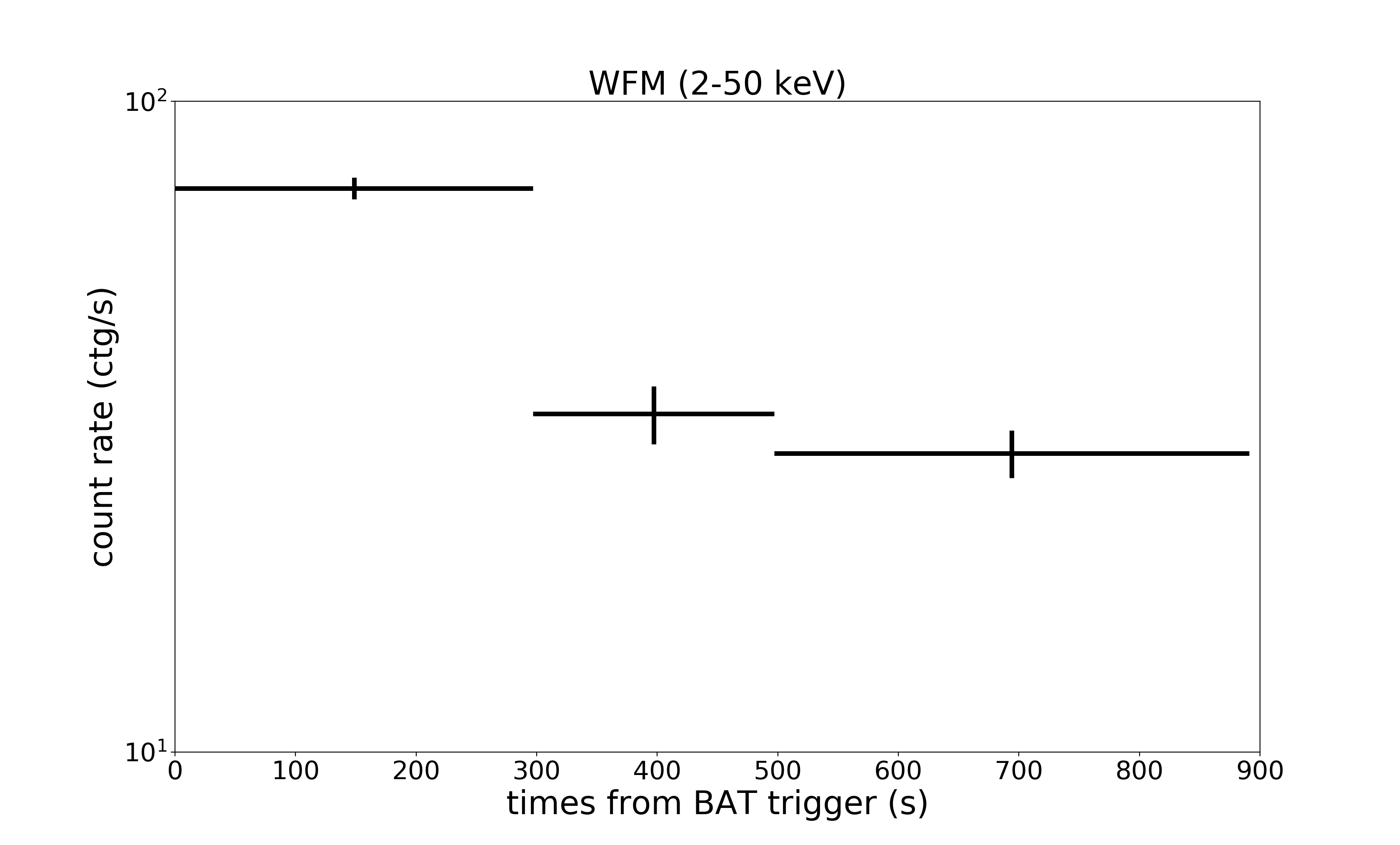}
\caption{Lightcurve of GRB 100316D as seen by Beppo-SAX, BATSE, INTEGRAL and the WFM-eXT according to our simulations.}
\label{fig:LightCurve_WFM}
\end{figure}

\begin{figure}
\centering
\includegraphics[scale=0.22]{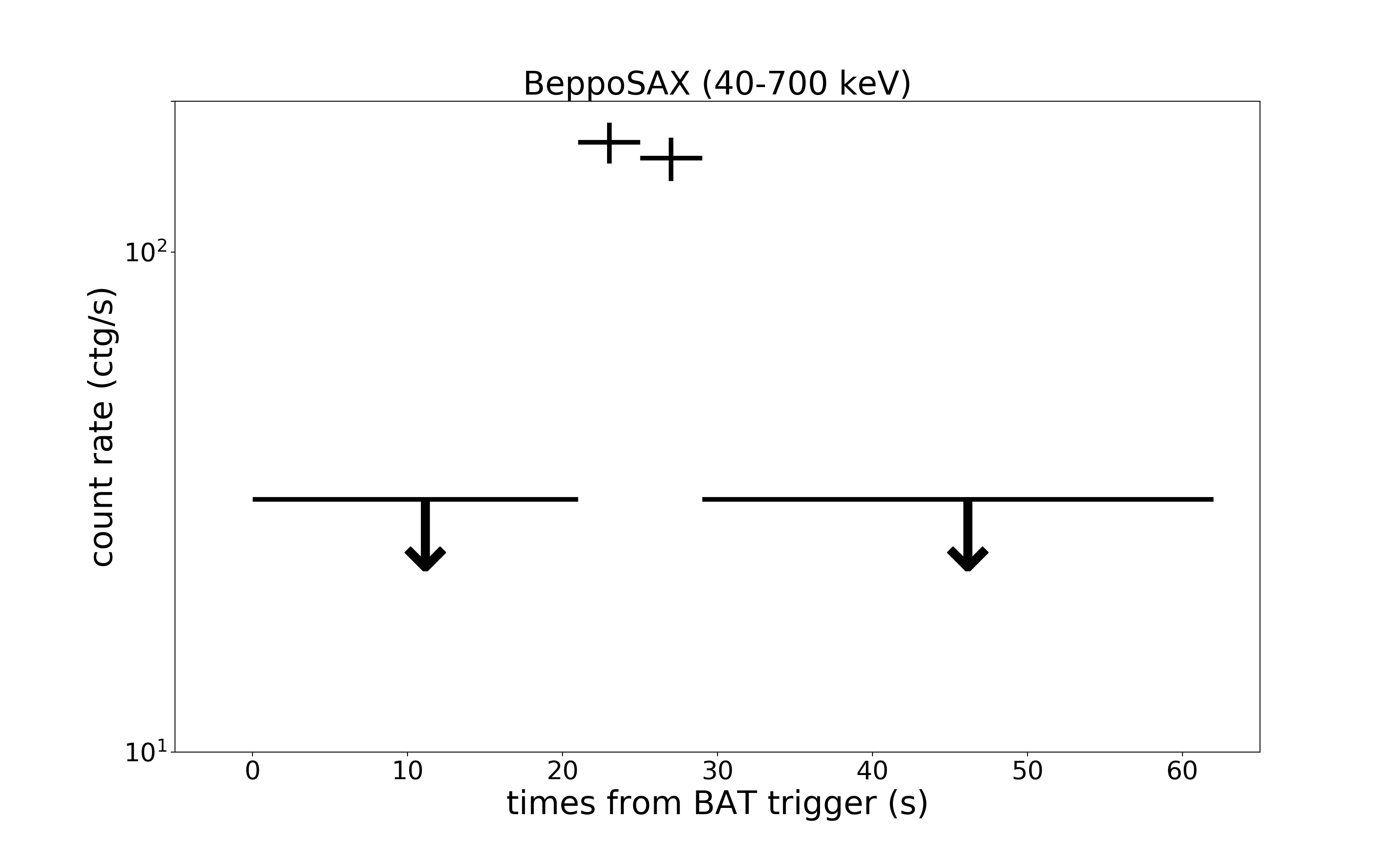}\\
\includegraphics[scale=0.22]{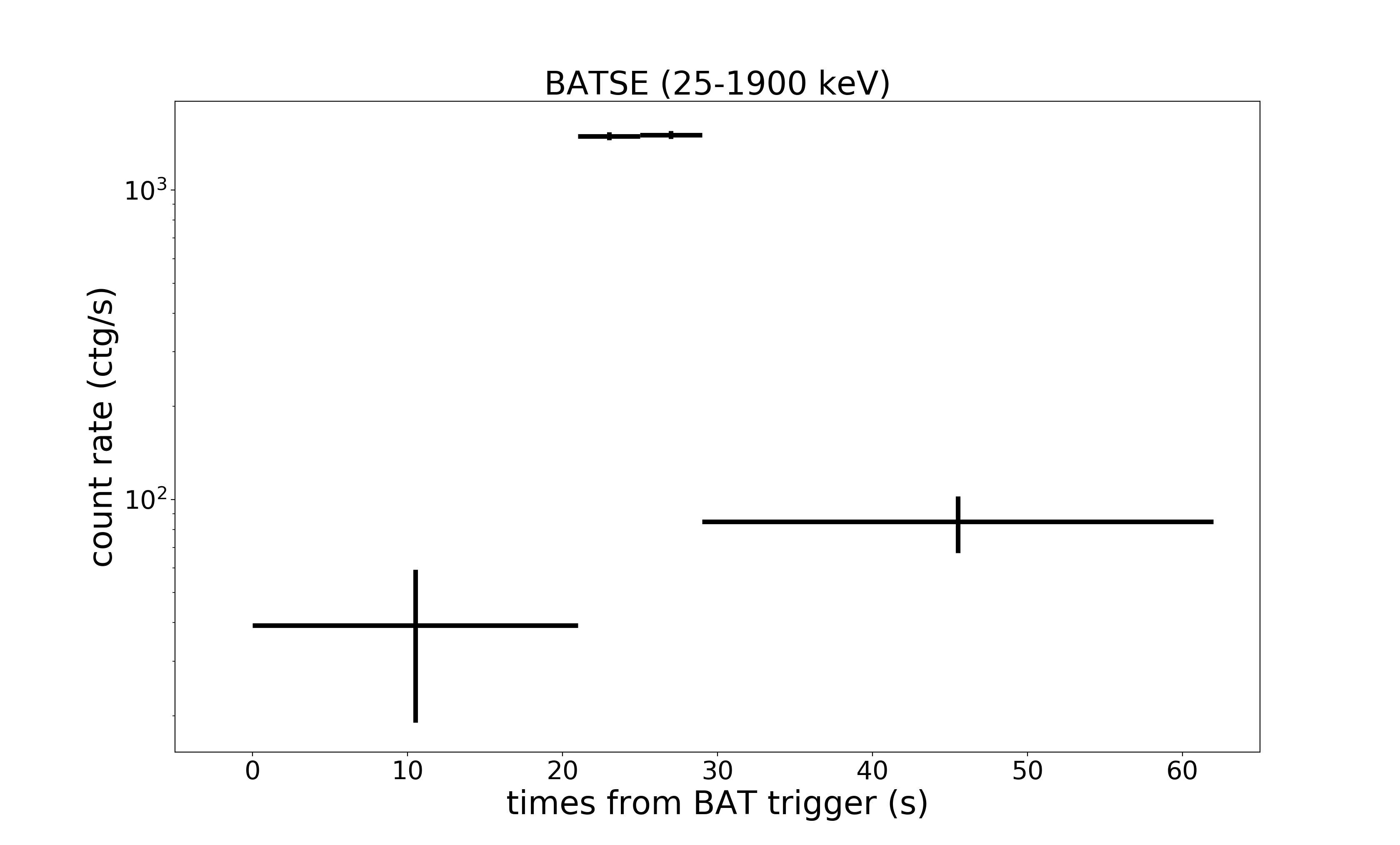}\\
\includegraphics[scale=0.22]{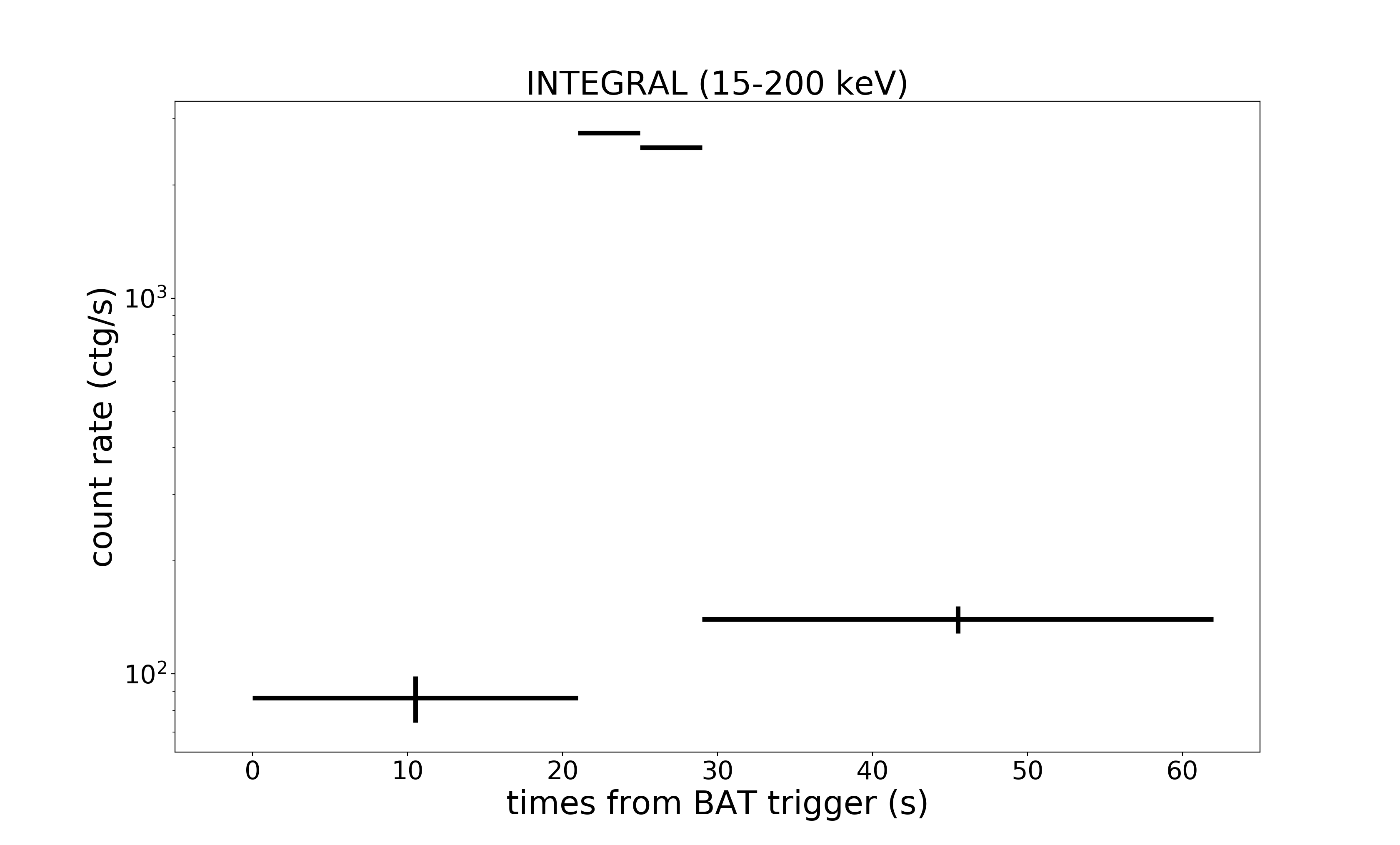}\\
\includegraphics[scale=0.22]{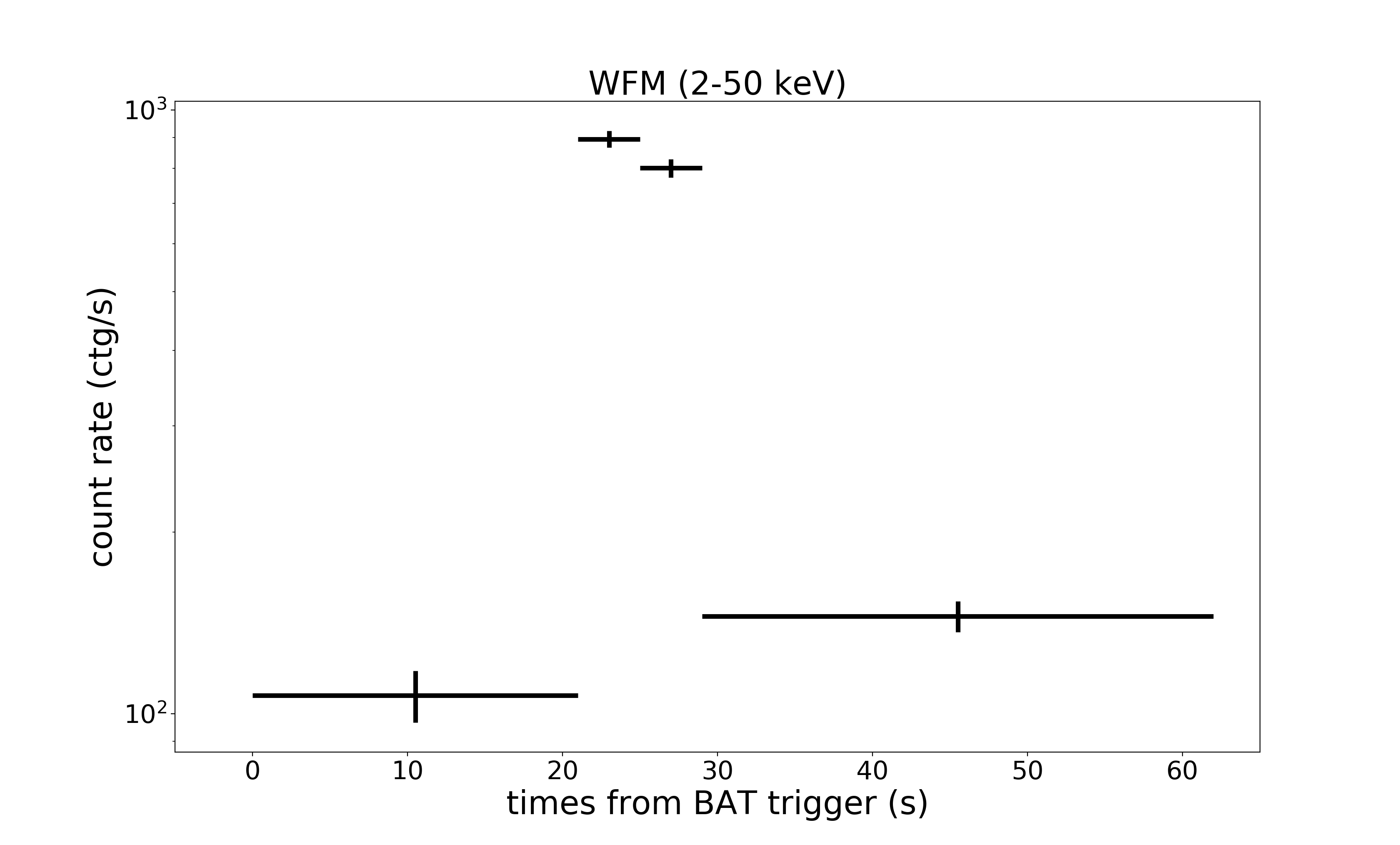}
\caption{Lightcurve of GRB 161219B as seen by Beppo-SAX, BATSE, INTEGRAL and the WFM-eXTP, according to our simulations.}
\label{fig:LightCurve_WFM}
\end{figure} 
\end{appendix}

\end{document}